\documentclass[preprint, authoryear]{elsarticle}

\biboptions{authoryear, semicolon}

\usepackage[utf8]{inputenc} 
\usepackage[T1]{fontenc}    
\usepackage{url}            
\usepackage{booktabs}       
\usepackage{amsfonts}       
\usepackage{nicefrac}       
\usepackage{microtype}      
\usepackage[colorlinks]{hyperref}

\usepackage{graphicx}
\usepackage{subcaption}

\usepackage{natbib}	

\usepackage[misc,geometry]{ifsym}

\usepackage{verbatim}
\usepackage{array}
\usepackage{color}
\usepackage{caption}
\usepackage{multirow}
\usepackage[table]{xcolor}

\newcolumntype{P}[1]{>{\centering\arraybackslash}p{#1}}

\renewcommand{\figureautorefname}{Figure }
\newcommand{\figuresautorefname}{Figures }
\renewcommand{\tableautorefname}{Table }
\renewcommand{\equationautorefname}{Equation }
\newcommand{\equationsautorefname}{Equations }

\renewcommand{\sectionautorefname}{Section }

\newcommand{\sectionsautorefname}{Sections }

\renewcommand{\vec}[1]{\mathbf{#1}}

\journal{Journal of Systems and Software}

\begin{document}

\hypersetup{
   linkcolor=red,
   filecolor=cyan,
   citecolor=blue,      
   urlcolor=magenta 
}

    
    \newcounter{DaveCommentCounter}
    \setcounter{DaveCommentCounter}{0}
    \newcommand{\dpt}[1]{
        \stepcounter{DaveCommentCounter}
        \textcolor{blue}{\emph{/* Dave's comment [\arabic{DaveCommentCounter}]: #1 */}}
    }

    \newcommand{\rrf}{
        \textcolor{blue}{\emph{\sc{/*need refs*/}}}
    }

\begin{frontmatter}

\title{Using Metamorphic Relations to Verify and Enhance Artcode Classification\tnoteref{t1}}
\tnotetext[t1]{This work is an extension of an MET '18 paper \citep{xu2018enhancing}, which was completed while the first author was a PhD student at University of Nottingham Ningbo China.}


\author[1]{Liming Xu}\ead{lx249@cam.ac.uk}
\author[2]{Dave Towey\corref{cor1}}\ead{dave.towey@nottingham.edu.cn}
\author[3]{Andrew P. French}\ead{andrew.p.french@nottingham.edu.cn}
\author[3]{Steve Benford}\ead{steve.benford@nottingham.edu.cn}
\author[4]{Zhi~Quan~Zhou}\ead{zhiquan@uow.edu.au}
\author[5]{Tsong Yueh Chen}\ead{tychen@swin.edu.cn}

\cortext[cor1]{Corresponding author}

\address[1]{Department of Engineering, University of Cambridge, Cambridge, United Kingdom}
\address[2]{School of Computer Science, University of Nottingham Ningbo China, Ningbo, China}
\address[3]{School of Computer Science, University of Nottingham, Nottingham, United Kingdom}
\address[4]{School of Computing and Information Technology, University of Wollongong, Australia}
\address[5]{Department of Computer Science and Software Engineering, Swinburne University of Technology, Australia}

\begin{abstract}
Software testing is often hindered where it is impossible or impractical to determine the correctness of the behaviour or output of the software under test (SUT), a situation known as the {\em oracle problem}. 
An example of an area facing the oracle problem is automatic image classification, using machine learning to classify an input image as one of a set of predefined classes. 
An approach to software testing that alleviates the oracle problem is metamorphic testing (MT). 
While traditional software testing examines the correctness of individual test cases, MT instead examines the relations amongst multiple executions of test cases and their outputs. 
These relations are called metamorphic relations (MRs): if an MR is found to be violated, then a fault must exist in the SUT. 
This paper examines the problem of classifying images containing visually hidden markers called {\em Artcodes}, and applies MT to verify and enhance the trained classifiers. 
This paper further examines two MRs, Separation and Occlusion, and reports on their capability in verifying the image classification using one-way analysis of variance (ANOVA) in conjunction with three other statistical analysis methods: t-test (for unequal variances), Kruskal-Wallis test, and Dunnett's test. 
In addition to our previously-studied classifier, that used Random Forests, we introduce a new classifier that uses a support vector machine, and present its MR-augmented version.
Experimental evaluations across a number of performance metrics show that the augmented classifiers can achieve better performance than non-augmented classifiers.  
This paper also analyses how the enhanced performance is obtained.
\end{abstract}


\end{frontmatter}


\section{Introduction}\label{sec:introduction}
Over the past two decades, machine learning techniques have been widely adopted by research communities (e.g., computer vision, bioinformatics, computational linguistics, and medical imaging) to solve a range of practical problems. 
For researchers in the machine learning and software testing communities, the ability to build accurate learning models and verify their quality is essential. 
Due to the nature of machine learning programs, test oracles (mechanisms to categorically determine if the software behaviour or output is correct) are generally very difficult to define. 
Hence, conventional software testing techniques may not be effective for detecting defects. 
The issue of how to ensure the quality of applications based on machine learning has become increasingly important \citep{xie2011testing}.

\textit{Metamorphic testing} (MT) is a testing technique that can alleviate the \textit{oracle problem} \citep{chen1998metamorphic,chen2003fault}, a major challenge in software testing. 
While conventional testing methods focus on verifying individual outputs, MT examines \textit{relations} among the inputs and outputs of multiple executions of the software under test (SUT). These \textit{relations} are called \textit{metamorphic relations} (MRs). Since the first MT paper \citep{chen1998metamorphic} was published in 1998, MT has been widely used to test software in various fields, including: 
scientific computing \citep{ding2016application}, numerical analysis \citep{chen2002metamorphic}, classification \citep{xie2011testing,xie2009application}, cybersecurity \citep{Tsong2016Cybersecurity}, image processing \citep{mayer2006random}, compilers \citep{le2014complier, Donaldson2017Automated}, search engines \citep{Zhou2016Search}, web security \citep{mai2020metamorphic}, and visualisation \citep{mcnutt2020surfacing}, among others. 
A body of literature also describes its integration with other testing techniques to improve their applicability and effectiveness. Comprehensive surveys about MT have also been recently published by \citet{segura2016survey} and \citet{chen2017review}. 

More recently, MT has been increasingly gaining interest in classic AI fields for testing systems powered by machine learning, including: 
machine translation \citep{Zhou2018Machine, He2019Structure}, autonomous driving \citep{zhang2018deeproad, zhou2019metamorphic}, and generic NLP (natural language processing) models \citep{ma2020metamorphic, marco2020beyond}. 
MT can have comparable bug-revealing effectiveness to model-based testing, and hence is a useful alternative to test an implementation, especially in situations where a model is expensive to construct \citep{hughes2020how}.

MT techniques have been used to test machine learning programs \citep{xie2011testing, xie2009application, murphy2008properties}. Machine learning techniques have also been used to automatically identify MRs, although so far only with simple MRs \citep{kanewala2013using}. \citet{xu2018enhancing} expanded the traditional role of MRs from software testing to a kind of {\it post adjustor} for a machine learning program, building a more accurate learning model using an example of the Artcode classification problem. Artcodes are visual codes whereby bespoke designs can be scanned to trigger the digital information attached to them. 
Artcodes may be disguised as normal images in the scene through their freeforms and complex aesthetic patterns --- and they may appear as any instances of semantic objects. 
Therefore, it is not straightforward for people to build {\it scan affordance} without the support of an alert system that can recognise the presence of Artcodes in the context. 
The core part of such an alert system is Artcode classification, which determines whether or not the Artcode-based augmented reality applications can work effectively.
We will present Artcode basics and Artcode classification in more detail in \sectionautorefname\ref{subsec:artcodesClassification}. More information about Artcode applications in augmented reality can be found in the literature, such as \citet{meese2013codes}, \citet{liming2017recognizing}, \citet{benford2018customizing}, and \citet{koleva2020designing}.

Two MRs, {\it Separation} and {\it Occlusion}, identified based on the category of the inputs, were introduced by \citet{xu2018enhancing}, who reported on their ability to improve the performance of the original classifier.  
Initial experimental evaluations showed that MRs could enhance the performance in this case of supervised Artcode classification.

In this paper, we further explore the Separation and Occlusion MRs, present more detailed experimental analyses, and generalise the ability of MRs in both verification and enhancement. Experiments were conducted to show not only the applicability of MT in verifying the correctness of the classifier, but also the improved performance obtained by the MR-augmented framework regardless of the chosen classification methods.
The new contributions of this paper are mainly threefold:  
\begin{enumerate}
	\item[1)] We report on the capability of the two MRs to verify the correctness of the previously introduced classification model \citep{liming2017recognizing} using a set of complementary statistical test methods.
	\item[2)] We analyse and discuss how the improved performance of the MR-augmented classifiers is achieved, explaining how the post adjustor rectifies incorrect predictions. 
	\item[3)] We introduce the use of a Support Vector Machine (SVM) as the classification algorithm in the original classifier and investigate its impact on the performance of the MR-augmented framework, comparing its performance with the MR-augmented classifier based on Random Forests (RF). 
\end{enumerate}

The rest of this paper is organised as follows. \sectionautorefname\ref{sec:preliminaries} gives a brief description of metamorphic testing and Artcode classification. \sectionautorefname\ref{sec:augmentedClassifier} presents the MR-augmented classification framework. 
The experimental studies examining the MRs' verification and enhancement capability are given in \sectionautorefname\ref{sec:experimentalEvaluation}.  
\sectionautorefname\ref{sec:discussion} analyses how the improved performance is obtained by MR augmentation. 
Finally, Section 6 concludes the paper, highlighting some areas for future work.

\section{Preliminaries}\label{sec:preliminaries}
\subsection{Metamorphic testing}\label{subsec:metamorphicTesting}
In software testing, a mechanism that can determine whether a test has passed or failed is called an {\it oracle}.
A situation where the oracle is not available, or is too expensive to be used, is known as the oracle problem \citep{BarrTSE15}. 
Metamorphic testing alleviates the oracle problem \citep{chen1998metamorphic}. It has been widely adopted in both academia and industry \citep{chen2003fault,liu2014how,lindvall2015metamorphic,segura2016survey, Zhou2016Search, Donaldson2017Automated,zhang2018deeproad,He2019Structure,mai2020metamorphic}. 
MT has successfully detected defects in mature software, including in extensively tested systems  \citep{ChenkuoToweyZhou:3RT:2015}.
A central part of MT is a set of MRs, which are relations among several related inputs and their corresponding outputs.  
While conventional testing approaches uncover software problems by examining the outcome of an individual input, MT detects the presence of a fault by cross-checking multiple related inputs and outputs with respect to MRs.

We next use a database management system (DBMS) example to illustrate the idea of MT.
Given two DBMS queries, such as the following: 
\begin{itemize}
    \setlength\itemsep{0px}
    \item[Q1:] {\em select} \textasteriskcentered ~{\em from} student {\em where} condition\_A {\em and} condition\_B;
    \item[Q2:] {\em select} \textasteriskcentered ~{\em from} student {\em where} condition\_B {\em and} condition\_A; 
\end{itemize}
the DBMS should return the same results
---
the outcome for a query with search conditions ``A'' and ``B'' and the query that swaps their order should be the same (which could represent an MR). 
Specifically, if the DBMS returns different results for the queries Q1 and Q2, then a fault must exist in the DBMS implementation.

As with all software testing, MT can only be used to check for the presence of bugs, not their absence \citep{dijkstra1970notes}. 
For example, a {\em faulty} DBMS implementation may somehow return the same results for queries Q1 and Q2:
thus, although violation of an MR means there must be some fault in the implementation, satisfaction of MRs cannot be taken to mean that the software is fault-free.
A key step in MT is the identification of appropriate MRs, which normally requires a good understanding of the problem domain.

\subsection{Artcode basics and classification}\label{subsec:artcodesClassification}
\begin{figure}[t]
        \centering
         \begin{subfigure}[b]{0.49\textwidth}
                \includegraphics[width=\textwidth]{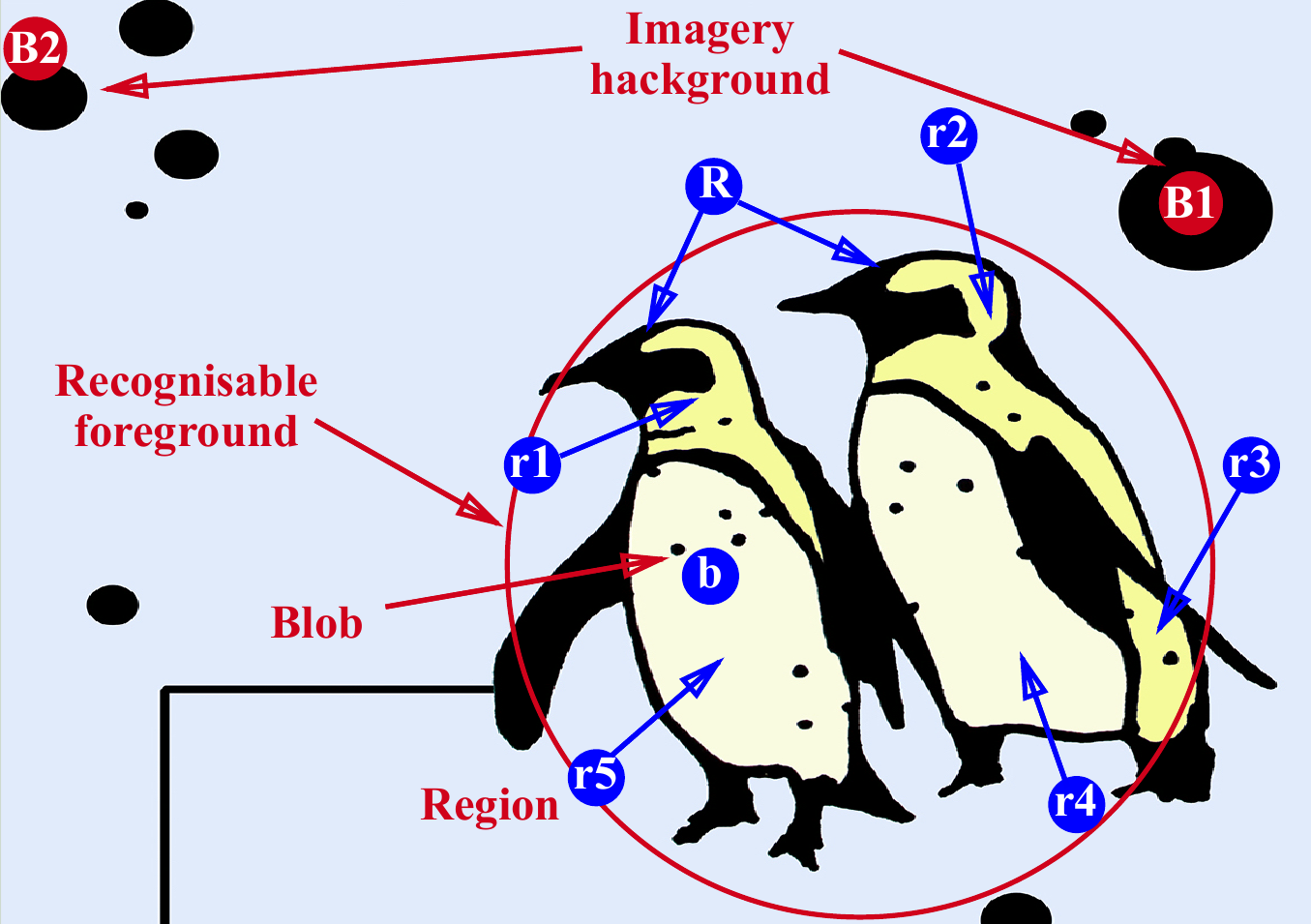}
                \caption{Artcode components illustration.}
                \label{fig:artcodeIllustration}
        \end{subfigure} 
        \begin{subfigure}[b]{0.49\textwidth}
                \includegraphics[width=\textwidth]{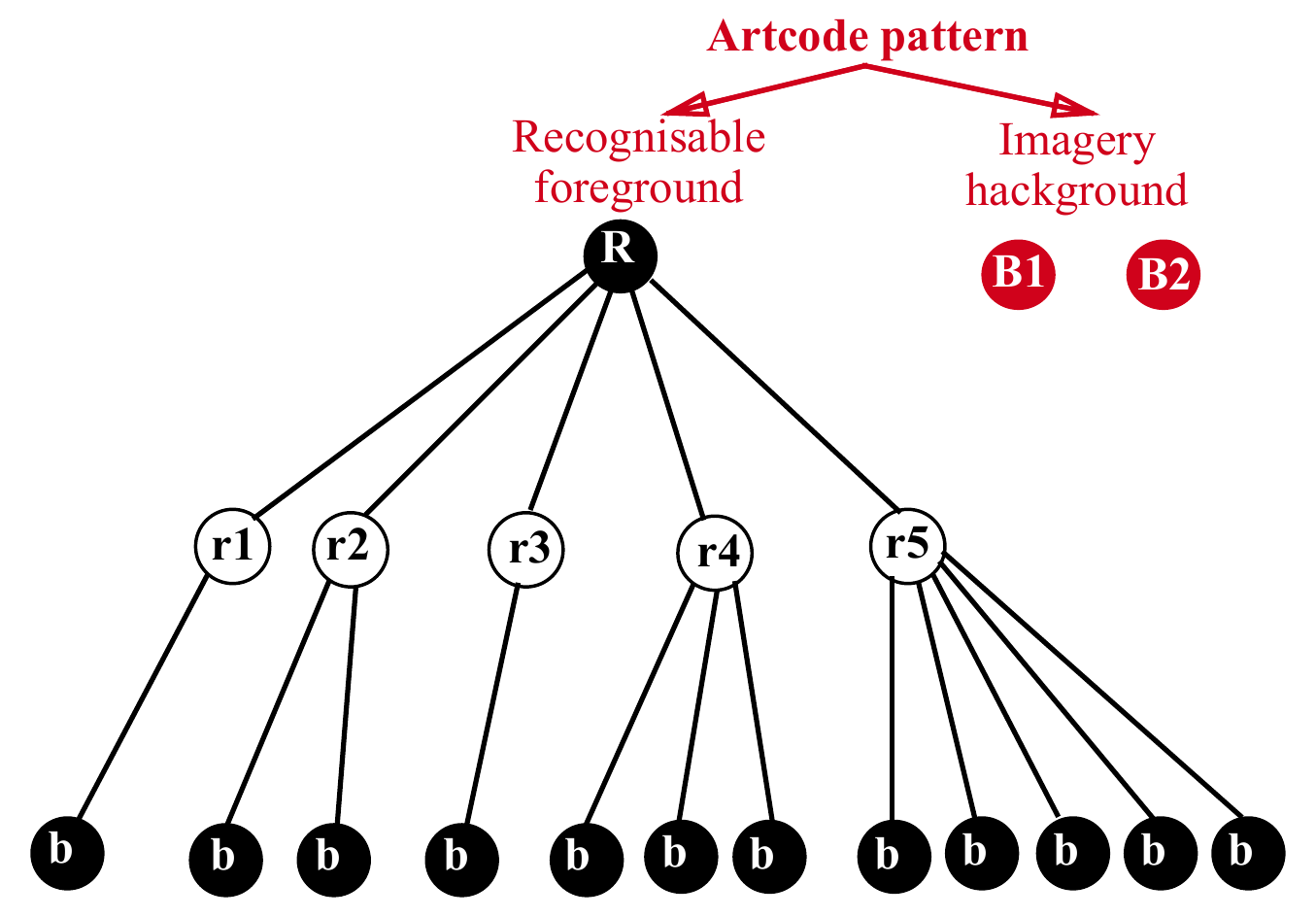}
                \caption{Region adjacency tree.}
                \label{fig:regionAdjacencyTree}
        \end{subfigure} 
         \caption{Illustration of the components of an Artcode (code: ``1-1-2-3-5'') and the region adjacency tree of its recognisable foreground.}\label{fig:artcodeBasics}
\end{figure}

Artcodes are human-designable topological visual markers that are both machine readable and meaningful to humans \citep{meese2013codes}.  
As illustrated in \figureautorefname \ref{fig:artcodeBasics}\subref{fig:artcodeIllustration}, a valid Artcode includes two parts: 
a recognisable foreground; and some background imagery. 
The recognisable foreground (the penguins annotated by the red circle) is a closed boundary that is split into several regions (usually five regions, annotated  {\tt r1} to {\tt r5} in \figureautorefname\ref{fig:artcodeBasics}\subref{fig:artcodeIllustration}), with each region containing one or more {\em blobs}
--- solid objects disconnected from the region edge. 
The numbers of these blobs in each region are sorted and joined with a separator to form a string of numbers, which can then be used to represent the Artcode. For example, the code for \figureautorefname \ref{fig:artcodeBasics}\subref{fig:artcodeIllustration} is ``1-1-2-3-5'', indicating that there are 1, 1, 2, 3, and 5 blobs found in the respective regions. 
Additionally, background imagery ({\tt B1} and {\tt B2} in \figureautorefname\ref{fig:artcodeBasics}\subref{fig:artcodeIllustration}) can be added, surrounding the recognisable foreground of an Artcode to enhance the aesthetics, but only if the background does not break the Artcode's topological structure \citep{costanza2009designable, meese2013codes}. For example, the black solid blobs around the penguins were {\em intentionally} added to enhance the beauty, but are unconnected to the actual code. 

The actual code of an Artcode is represented by a region adjacency tree (RAT) \citep{Costanza2003}; 
the RATs of the recognisable part of the penguin Artcode and the two background elements are shown in \figureautorefname \ref{fig:artcodeBasics}\subref{fig:regionAdjacencyTree}.
According to the Artcode system, the components are the sets of pixels that are connected to each other, and are known as {\em connected components}. 
These connected components are referred to as: {\em root boundary}, {\em region}, {\em blob}, and {\em background imagery}, depending on their use in the Artcode's context. 
The root boundary ({\tt R}) contains several holes (regions) with each having a number of connected components without holes. The number of components is determined by the containment relationship rather than geometrical shapes, as shown in \figureautorefname \ref{fig:artcodeBasics}. 
The components can be any shapes, and this freeform property 
--- 
with little restriction on shapes 
--- 
can be an opportunity for designers to create aesthetic, interactive graphics. This property allows Artcode objects to look like an instance of any semantic object classes --- animals, flowers, and fish can be recognised as Artcodes if they are designed according to Artcode drawing rules (see \figureautorefname \ref{fig:artcodesExamples}).

{\em Redundancy} is allowed in Artcode design --- multiple Artcodes with the same topology but different geometry can appear in an Artcode. 
Artcodes have been explored in a wide range of contexts \citep{meese2013codes, benford2015augmenting, benford2015carolan, ng2016design, thorn2016exploring, benford2016accountable, preston2017enabling, benford2018customizing, koleva2020designing} since \citet{costanza2009designable} first proposed D-touch markers, whose drawing rules the Artcode system implements and extends.

\begin{figure}[t]
        \centering
         \begin{subfigure}[b]{0.49\textwidth}
                \includegraphics[width=\textwidth]{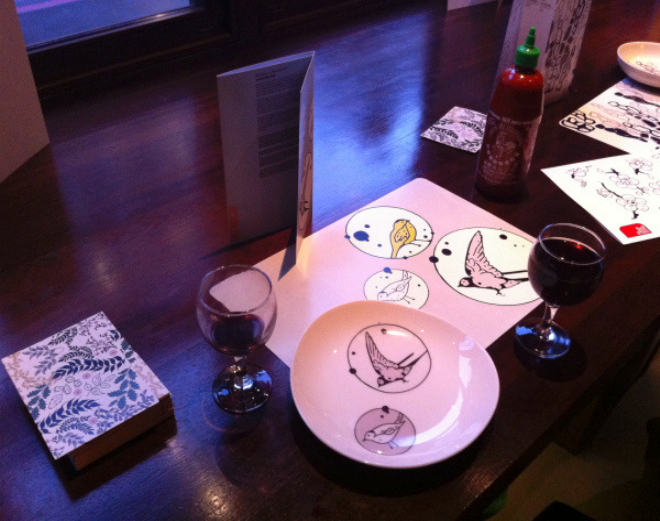}
                \caption{Artcodes-decorated dining context}
                \label{fig:artcodesResContext-1}
        \end{subfigure} 
        \begin{subfigure}[b]{0.49\textwidth}
                \includegraphics[width=\textwidth]{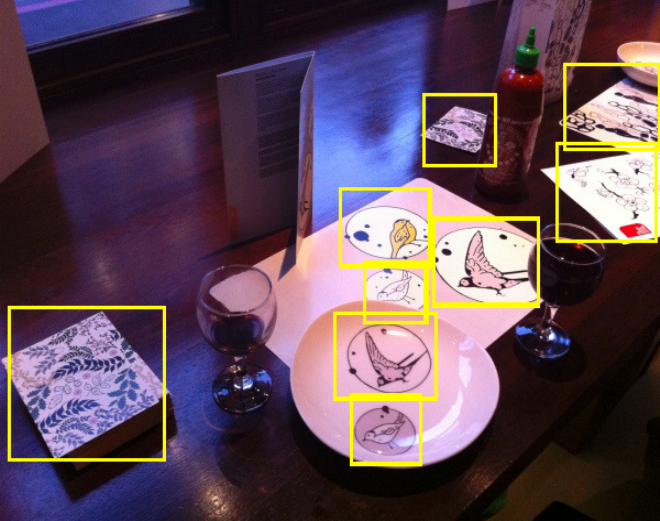}
                \caption{Artcode examples detected}
                \label{fig:artcodesResContext-2}
        \end{subfigure} 
         \caption{Illustration of Artcode detection.}\label{fig:artcodesResContext}
\end{figure}

\paragraph{Artcode classification}
\figureautorefname\ref{fig:artcodesResContext}\subref{fig:artcodesResContext-1} shows Artcodes being used to augment a dining context, in which the surfaces of objects (menu, plate, and mat) are decorated with Artcodes. In order to alert people to the presence of Artcodes before triggering their further decoding, the first step is to determine whether or not an input image or an image patch contains Artcodes (see \figureautorefname\ref{fig:artcodesResContext}\subref{fig:artcodesResContext-2}). 
This step involves classification which determines whether an input image is an Artcode or not. 
This task of {\em Artcode classification}\footnote{{\em Artcode classification} and {\em Artcode detection} are used interchangeably in this paper.} involves classifying an input image as either containing an Artcode or not, labelled {\em Artcode} or {\em non-Artcode}. There is, visually, no obvious difference in appearance or geometrical shape between the two classes (see the examples in \figuresautorefname\ref{fig:nonartcodesExamples} and \ref{fig:artcodesExamples}). 
The geometrical freeform property differentiates Artcodes from other well-known markers, such as barcodes \citep{woodland1952classifying}, QR codes \citep{wave2015information}, ARTags \citep{Fiala2005}, or RUNE-tags \citep{bergamasco2011rune}. 
Artcodes, as a type of augmented reality technique, have been adopted in many situations (as described in \sectionautorefname \ref{subsec:artcodesClassification}) {\color{blue}to augment} the meanings of the objects in aesthetic-centred contexts. The triggering of the digital information depends on whether or not the presence of Artcodes in the scene is recognised; therefore, Artcode classification is vital for the correct use of Artcode applications and can provide guidelines for other visual codes-based augmented reality techniques. 
More information about Artcode basics and classification can be found in work such as \citet{costanza2009designable} and \citet{liming2017recognizing}.

\section{MR-augmented classification framework}\label{sec:augmentedClassifier}
\begin{figure}[t]
\centering
\includegraphics[width=\textwidth]{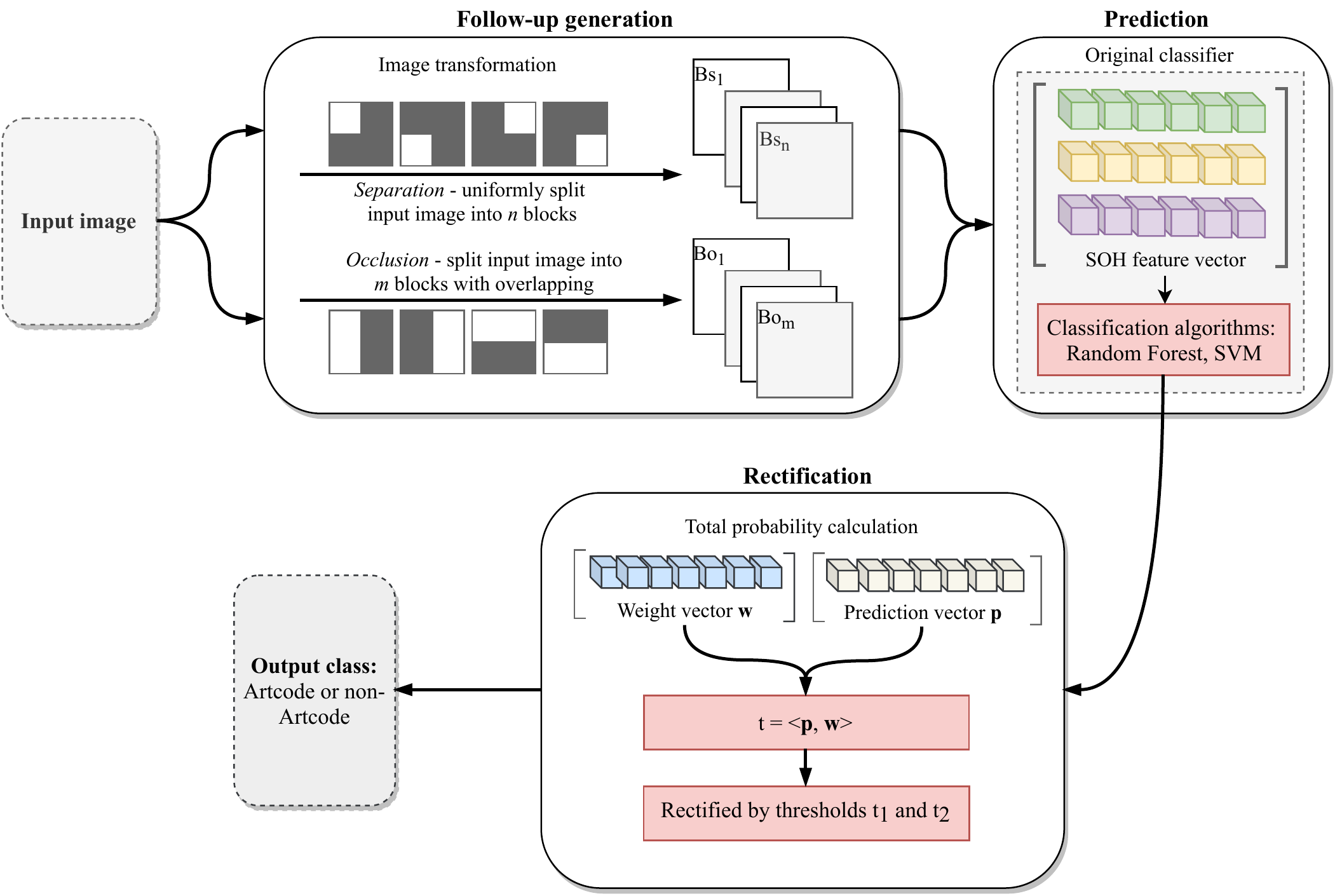}
\caption{MR-augmented classification framework. The framework includes three stages: Follow-up generation, Prediction, and Rectification.}\label{fig:augmentedClassifierFramework}
\end{figure}

Conventional classification typically involves two steps: first, create feature vectors that {\em distinctively} represent each class; and, second, train classification algorithms to predict the class of individual inputs. 
\citet{xu2018enhancing} proposed two MRs, Separation and Occlusion, through examination of the differences in aggregated probability of image blocks being classified as {\em Artcode} between the Artcode and non-Artcode class. 
The two MRs were then used to enhance the classifier's performance based on conventional classification methods by adding a step before and after classification by this {\em base} classifier (referred to as the {\em original} classifier), resulting in an {\em MR-enhanced classifier}.

We have refined the MR-augmented classifier previously proposed in \citet{xu2018enhancing} to present a new use of MRs in the verification of its correctness. 
As shown in \figureautorefname \ref{fig:augmentedClassifierFramework}, the MR-augmented classifier framework includes three stages: Follow-up generation, Prediction, and Rectification. 
Follow-up generation involves building inputs for the prediction stage using MR-defined image transformations. 
The second stage makes predictions about these inputs using commonly-used classification models. 
The third stage may adjust or rectify the results generated in the prediction stage. 
These three stages are described in detail in \sectionsautorefname \ref{subsec:metaRelations} to  \ref{subsec:rectification}.

\begin{figure}[t]
        \centering
        \begin{subfigure}[b]{0.35\textwidth}
                \includegraphics[width=\textwidth]{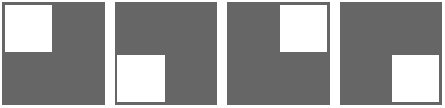}
                \caption{Separation masks}
                \label{fig:separationMasks}
        \end{subfigure} 
        \hspace{3pt}
        \begin{subfigure}[b]{0.35\textwidth}
                \includegraphics[width=\textwidth]{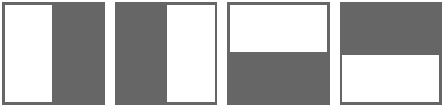}
                \caption{Occlusion masks}
                \label{fig:occlusionMasks}
        \end{subfigure} 
        \caption{Separation and occlusion masks.}
        \label{fig:separationOcclusionMasks}
\end{figure}

\subsection{Follow-up generation}\label{subsec:metaRelations}
The core activity of the follow-up generation stage is to identify MRs and to construct the inputs based on the defined MRs. 
The identification of MRs in image classification is often done by examining the different image transformations, such as translation, rotation, and scaling. 
Based on the observation that the image blocks of Artcode images are more likely to be classified as {\em Artcode} than the blocks of non-Artcodes, two MRs, Separation and Occlusion, were proposed, using straightforward image operations: {\em uniform} and {\em non-uniform} separation. 
This stage accepts an entire image as input, and outputs image blocks generated from the operations defined by the two MRs.

\subsubsection{Separation MR}\label{subsec:mrSepration}
Separation involves splitting the input image uniformly into a number of sections, or {\em blocks}. 
For example, \figureautorefname \ref{fig:separationOcclusionMasks}(\subref{fig:separationMasks}) shows separation masks to generate four uniform blocks by intersecting them with input images. 
This MR is based on the observation that the blocks of Artcode images would be predicted to be Artcode with a higher {\it likelihood} than the blocks of non-Artcode images. If the number of blocks is appropriately selected, this difference in the aggregated likelihood (probability) of all blocks may provide clues for classification. 
The Separation MR can be formulated as: 
\begin{equation}\label{eq:separation}
	\sum_{i=1}^n{\Pr(B^{i}_{s_a})} \ge \sum_{i=1}^n{\Pr(B^{i}_{s_n})}
\end{equation}
where $n$ is the number of image blocks;  
$\Pr(\mathord{\cdot})$ is the probability for it to be classified as an Artcode by the original classifier; 
and $B^{i}_{s_a}$ and $B^{i}_{s_n}$ denote the $i$th block of the Artcode and non-Artcode images  after separation, respectively.

\subsubsection{Occlusion MR}\label{subsec:mrOcclusion}
Occlusion is similar to separation, except that the image blocks are not split uniformly 
--- 
overlapping among image blocks is permitted. 
As shown in \figureautorefname \ref{fig:separationOcclusionMasks}\subref{fig:occlusionMasks}, four occlusion masks are provided to intersect with the input image, so that the image blocks outlined by the white regions will be generated. 
Occluded Artcode images generally preserve the properties of the input Artcode images --- 
half of an Artcode image usually has a higher likelihood of being classified as Artcode by the original classifier than a quarter of the image; 
this property may not be preserved for non-Artcode images: 
occluded non-Artcode images may have the equivalent likelihood as the entire non-Artcode images of being predicted as non-Artcode. 
Based on this observation, MR Occlusion can be formulated as:
\begin{equation}\label{eq:occlusion1}
	\sum_{i=1}^m{\Pr(B^i_{o_a})} \ge \sum_{i=1}^m{\Pr(B^i_{o_n})}
\end{equation}
\begin{equation}\label{eq:occlusion2}
	\Pr(B^i_{o_a}) \ge \Pr(B^i_{s_a}) 
\end{equation}
where $m$ is the number of masks; 
$B^i_{o_a} = \cap(I_a, M_i)$ and $B^i_{o_n} = \cap(I_n, M_i)$ outputs the overlapping areas of Artcode and non-Artcode images, $I_a$ and $I_n$, and the $i$th mask $M_i$; and $B^{i}_{o_a}$ and $B^{i}_{o_n}$ denote the $i$th block of the Artcode and non-Artcode images generated after occlusion, respectively.

The Separation and Occlusion MRs are both processed by comparing the aggregated likelihood of predicting the generated image blocks with the probability of predicting the entire input image. 
They are based on the observation that the topological structure of an Artcode image, as a global property, may be preserved, even after splitting. 
Uniform separation with (separation) and without (occlusion) overlapping enable the generated image blocks to cover the possible distribution of Artcodes in an image, especially considering their freeform geometric shapes. 
In addition, the masks with varying sizes can adapt to the Artcodes' scales. 
Therefore, they complement each other, and are combined together to obtain a better augmentation performance.

\subsection{Prediction}\label{subsec:stagePrediction}
In order to predict the class of an input image or block, a classification model that includes feature vector and classification algorithms (using random forests or support vector machines) needs to be built. 
The Artcode classification model is built using the Shape of Orientation Histograms (SOH) feature vector \citep{liming2017recognizing}, which was specially designed for describing topological visual markers such as Artcodes. An SOH is constructed based on the {\em translational symmetry} and {\em smoothness} of the orientation histogram, which is a feature vector developed by \citet{mcconnell1986method} for pattern analysis in both static and dynamic modes, and was adopted by \citet{freeman1991design} for recognising hand gestures.

Instead of describing the local geometry or structure, an SOH describes Artcodes by representing their  topological structure. 
As previously reported \citep{liming2017recognizing}, the orientation histogram of an Artcode displays  horizontally translational symmetry, and is relatively smoother than that of a non-Artcode. 
The SOH is then constructed by quantifying these two aspects of the orientation histogram of the input images using similarity measurements, such as Procrustes distance \citep{moser1965volume} and $\chi^2$ distance \citep{greenacre2017correspondence}. 
When all images are represented by their respective SOH vectors, classification algorithms using random forests or SVM are trained and used to predict the classes of the input images. 
The output of the prediction stage is a vector of labels of the input image blocks fed by the follow-up generation stage. This vector is referred to as the {\em prediction vector} $\vec{p}$.

\subsection{Rectification}\label{subsec:rectification}
Unlike most deterministic software, classification is based on {\em statistics}, or is learned from past experience. 
Given an input, the output of a classifier is a {\em probabilistic} classification of belonging to a predefined class. 
In other words, before execution of the classifier, only the likelihood of the input being classified as a class or not is known beforehand. Therefore, in order to enable incorporation of the MRs described above, an augmented classifier integrating the MRs was designed based on {\em probability}, adding an adjustor (or rectifier) to the conventional classification pipeline \citep{xu2018enhancing}.

As defined in \equationsautorefname \ref{eq:separation} to \ref{eq:occlusion2}, the likelihood of image patches belonging to the two classes, generated in the follow-up generation stage, may be different. Therefore, a weight vector that contains different \textit{weight} (i.e., likelihood) values is assigned to them. This vector, which has same dimensionality as the prediction vector $\vec{p}$, is referred to as the \textit{weight vector} $\vec{w}$.

Given a prediction vector $\vec{p}=\left(p_{s_1}, \dots ,p_{s_n}, p_{o_1},\dots, p_{o_m}\right)$, and a weight vector $\vec{w}=\left(w_{s_1}, \dots ,w_{s_n}, w_{o_1},\dots, w_{o_m}\right)$ 
--- 
where $p_i$ is the predicted class of the $i$th image patch by the original classifier; 
$w_i$ is the weight assigned to the $i$th image patch 
(which is, in fact, the weight of the separation or occlusion mask); 
and $n$ and $m$ are the numbers of image patches generated by the two MRs 
--- 
the inner product of $\vec{p}$ and $\vec{w}$ is the {\em aggregated likelihood} of belonging to the Artcode class ($\mathrm{\rho}$-value), which is defined as:
\begin{equation}\label{eq:aggregatedLikelihood}
\vec{\rho} = \langle \vec{p}, \vec{w}\rangle = \Big(\sum_{i=1}^{n}p_{s_i}\cdot w_{s_i} + \sum_{i=1}^{m}p_{o_i}\cdot w_{o_i}\Big)
\end{equation}
The aggregated likelihood is also known as the {\em total probability} \citep{xu2018enhancing, liming2020thesis}. The augmented classifier predicts the label of the input by comparing the $\rho$-value with the given thresholds $t_1$ and $t_2$, using the following decision rules:
if $\mathrm{\rho} < t_1$, then it is a non-Artcode;
if $\mathrm{\rho} \ge t_2$, then it is an Artcode; 
otherwise, the input retains the original classifier's prediction result.

\section{Experimental studies}\label{sec:experimentalEvaluation}
This section presents the experimental study, including the evaluation dataset and the set-up of the experiment. 
The experimental results of verifying and enhancing the original classifier, and the performance comparison between the RF-based and SVM-based classifiers, are also described in this section.

\begin{figure}[t]
        \centering
         \begin{subfigure}[b]{\textwidth}
                \includegraphics[width=\textwidth]{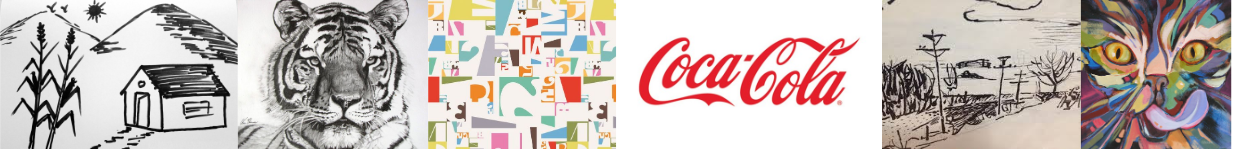}
        \end{subfigure} 
        \caption{Non-Artcode examples selected from the Artcode dataset.}\label{fig:nonartcodesExamples}
\end{figure}
\begin{figure}[t]
        \centering  
        \begin{subfigure}[b]{\textwidth}
                \includegraphics[width=\textwidth]{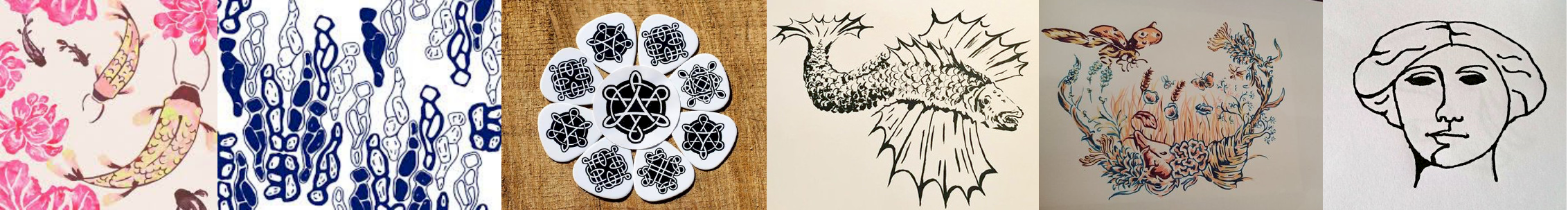}
        \end{subfigure} 
        \caption{Artcode examples selected from the Artcode dataset. Artcodes are visually ``hidden'' or even ``invisible'' markers. Similar to barcodes and QR codes, they can be scanned to trigger the digital information attached within. The code embedded in an Artcode is a string of numbers of {\em blobs} in each ``hollow'' {\em region}. For example, the code of the 6th image is ``1-1-1-1-2''.}\label{fig:artcodesExamples}
\end{figure}

\subsection{Dataset} \label{subsec:dataset}
In order to study the Artcode classification problem, a dataset containing 47 Artcode and 116 non-Artcode images was used for experimental study. 
To the best of our knowledge, this is the first dataset available for studying Artcode classification.
The non-Artcode images (including logos, drawings, advertisements, paintings, and graphics) were all created by humans, and were intentionally selected such that they would appear very similar to actual Artcode images \citep{xu2018enhancing}. 
This means that the dataset is very challenging for Artcode classification. 
As shown in \figuresautorefname\ref{fig:nonartcodesExamples} and \ref{fig:artcodesExamples}, Artcode examples look very similar to the non-Artcode images, which can make it very difficult to distinguish between the two classes through visual inspection alone. 
Because Artcodes are manually created by designers, the number of available Artcodes is currently small and slightly imbalanced, but work is ongoing to extend the dataset\footnote{\url{https://www.artcodes.co.uk/creations/}}. However, it is not possible to create hundreds of Artcode samples within a short time frame, much less increase the number to thousands or millions, like other common image classification tasks. Rather than devoting the very large effort necessary to expand the size of the dataset, we accepted this situation (of a small, imbalanced dataset), and adopted measures to address it, and mediate its impact: 1) We used classification methods that are effective on small datasets; 2) we adopted a group of carefully-considered performance evaluation metrics that are capable of evaluating classifiers used on imbalanced datasets; 3) we employed cross-validation techniques for experimental evaluation; and 4) we applied appropriate statistical methods to verify whether or not the improved performance was indeed attributable to the MR augmentation.

\subsection{Cross-validation}\label{crossValidation}
Cross-validation is a commonly-used model validation technique for assessing how a learning model will generalise to a dataset \citep{kohavi1995study, devijver1982pattern, seni2010ensemble}. 
A major reason for using cross-validation, rather than using the conventional validation method that partitions the dataset into two sets (70\% for training and 30\% for testing), is that sufficient data may not be available for training and testing the model without compromising its generalisation and prediction capability.

Considering the limited number of samples in the Artcode dataset, a 5-fold  cross-validation was used to ensure sufficient training and testing set sizes for performance evaluation. 
A $k$-fold cross-validation involves randomly partitioning a dataset into $k$ equally-sized subsets, keeping one subset as validation data for testing the trained model, and using the remaining $k-1$ subsets as training data. The process is then repeated $k$ times (the \textit{folds)}.

\subsection{Study 1 -- Verification}\label{subsec:verificationExperiment}
MT attempts to verify the software through examination of whether or not the identified MRs are violated:
as explained in \sectionautorefname\ref{subsec:metamorphicTesting}, violation of the Separation and/or Occlusion MR would indicate that the original classifier has not been correctly implemented.

Due to the {\em uncertainty} of a prediction by the original classifier, we explored its correctness by examining the weighted sum of probability of all image blocks of an input image being classified as {\em Artcode}  
--- the aggregated likelihood $\rho$ --- seeing if Artcodes and non-Artcodes had significant differences in the aggregated likelihood. 
Given input groups of $N$ Artcode and $M$ non-Artcode images, after the follow-up generation and prediction stages (\figureautorefname\ref{fig:augmentedClassifierFramework}), the two classes then have two sets of $\rho$-values calculated based on \equationautorefname\ref{eq:aggregatedLikelihood}:
\begin{equation}
 \rho_{G_a} = \big \{ \rho_{a_i} \mid  i = 1, \dots, N \big \}, \;\;
 \rho_{G_n} = \big \{ \rho_{n_i} \mid i = 1, \dots, M \big \} 
\end{equation}
where $\rho_{G_a}$ and $\rho_{G_n}$ denote the sets of aggregated likelihood of image samples of Artcode ($G_a$) and non-Artcode ($G_n$) category, respectively.

We then examined the implementation correctness by checking whether or not the relationship that $\rho_{G_a}$ and $\rho_{G_n}$ are significantly different was violated. 
Because of the probabilistic nature of the classifier, we used one-way analysis of variance (ANOVA) to assess the possible violation. 
ANOVA is a form of statistical hypothesis-testing that can be used to analyse whether or not there are statistically significant differences among the means of independent groups. 
We used ANOVA to examine if there was a statistically significant difference between the two groups $\rho_{G_a}$ and $\rho_{G_n}$ 
--- 
overall, the $\rho_{G_a}$ may be {\em significantly} ``greater'' than $\rho_{G_n}$ from a statistical perspective 
--- 
using separation and occlusion. 
If not, the classifier may be incorrectly implemented.
When employing one-way ANOVA, it is assumed that the variances of different groups are equal and that the $\rho$-values are normally distributed. 
However, although the two groups were independently selected and members in groups $G_{n_i}$ were randomly selected, it was not certain that the {\em normality} and {\em equal variance} assumptions were satisfied in the experiment. 
Although one-way ANOVA is not very sensitive to deviations from normality, according to simulation results by \citet[pp. 157--164]{mcdonald2009handbook}, we conducted further studies to consider situations of {\em non-normality} and 
{\em unequal variances}. 
In contrast to examining if the two assumptions were satisfied, we consolidated the experiment by introducing two more statistical test methods: 
t-test (for unequal variances)
---
which can be used to determine if the means of two groups $\rho_{G_n}$ and $\rho_{G_{n_i}}$ are significantly different when the variances are unequal; and 
Kruskal-Wallis test \citep{kruskal1952use} (also called one-way ANOVA on ranks, denoted  ANOVA\_ranks)
--- 
which is suitable for studying the difference between the means of two groups under non-normality situations. 
Hence, one-way ANOVA in conjunction with t-test and ANOVA\_ranks can effectively evaluate the difference between the mean $\rho$-values of the two groups under the aforementioned situations. 
As the comparisons between $\rho_{G_a}$ and each $\rho_{G_n}$ using these three methods were conducted separately, rather than simultaneously, we also used Dunnett's test \citep{dunnett1955multiple} as a {\em post hoc test} method. 
Dunnett's test is a {\em multiple comparison} procedure that enables one-to-many comparisons {\em simultaneously} to check if significant differences exist between the Artcode group $\rho_{G_a}$ and each of the  non-Artcode groups $\rho_{G_{n_i}}$.
The following sections present this verification examination, including detailing the experimental setting and results.

\subsubsection{Experimental setting}\label{subsec:verificationExperimentalSetting}
In order to examine the correctness of the classifier, we checked for violation of the MRs through examination of the variation of $\rho$-values between the two classes. 
Considering the different sizes of $G_a$ and $G_n$ ($N < M$) 
--- 
$G_n$ is considerably larger than $G_a$ 
--- 
$N$ elements were randomly selected from $G_n$ each time, with this process run $K$ times to generate $K$ non-Artcode groups $G_{n_i}, i = 1\,\dots\,K$. 
One-way ANOVA, t-test (for unequal variances), one-way ANOVA on ranks, and Dunnett's test were conducted to examine if there was a significant difference between $\rho_{G_a}$ and each  $\rho_{G_{n_i}}$. 
To reduce variance, we randomly selected $K$ groups, $G_{n_i}, i = 1\,\dots\,K$, from the non-Artcode group $G_n$, in which each $G_{n_i}$ had the same size as the group $G_a$. 
We used the RF-based original classifier as the SUT for study, and a 5-{\em fold} cross-validation  to obtain the prediction results of the image blocks generated by the follow-up generation stage. 
The weights of $w_{s_i}$ ($i = 1\,\dots\,n$) and $w_{o_j}$ ($j = 1\,\cdots\, m$)
were all assigned the same values, meaning that all image blocks generated based on separation or occlusion had the same weights 
--- 
having the same likelihood to contain Artcodes.
The weights between the images blocks for separation may be different from those for occlusion.

\begin{figure}[t!]
	\centering
    \begin{subfigure}[b]{\textwidth}
        \includegraphics[width=1\textwidth]{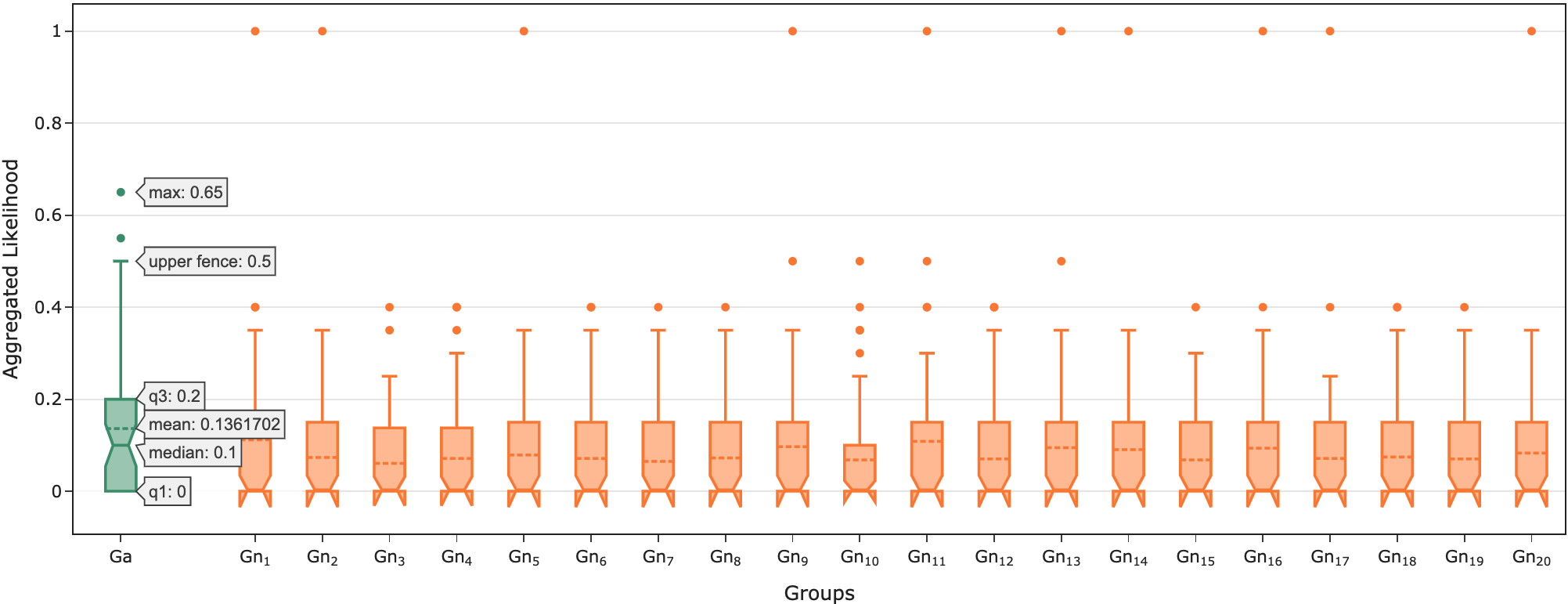}
    \end{subfigure}
    \caption{Boxplot of the aggregated likelihood ($\rho$) of Artcode group (${G_a}$) and non-Artcode groups ($G_{n_{1-20}}$). The dashed line in each box denotes the mean aggregate likelihood of the group, i.e., $\overline{\rho_{G}}$. The grey arrowed box annotations show the mean, maximum, minimum, median, first quartile (q1) and third quartile (q3) of the Artcode group.}
    \label{fig:groups-boxplot}
\end{figure}

\subsubsection{Results}\label{subsec:verificationResults}
\figureautorefname \ref{fig:groups-boxplot} presents a boxplot of the aggregated likelihoods of the group $G_a$ and $G_{n_{1-20}}$; and
\tableautorefname \ref{tbl:verificationResults} shows the $\mathrm{p}$-values for comparisons between $\rho_{G_a}$ and each $\rho_{G_{n_i}}$, according to the four tests. 
The average aggregated likelihoods (dashed line in \figureautorefname \ref{fig:groups-boxplot}) of all images in Artcode and non-Artcode categories were calculated using the following formula: 
\begin{equation}
	\overline{\rho_{G}} = \frac{1}{N}\sum_{i=1}^N\rho_i
\end{equation}
where $\overline{\rho_G}$ is the mean aggregated likelihood of group $G$. 
The mean aggregated likelihood of all randomly generated non-Artcode groups $G_{n_i}$ is defined as:
\begin{equation}
	\overline{\rho_{G_n}} = \frac{1}{K}\sum_{i=1}^{K}\overline{\rho_{G_{n_i}}}
\end{equation}
The mean aggregated likelihood of all groups is defined as: 
\begin{equation}
	\overline{\rho_{G_a}+\rho_{G_n}} = \frac{1}{K+1}\big (\sum_{i=1}^{K}\overline{\rho_{G_{n_i}}} + \overline{\rho_{G_a}} \big )
\end{equation}
where $K$ is the number of groups randomly selected from $G_n$; and 
$N$ and $M$ are the total number of Artcode and non-Artcode images in the Artcode dataset, respectively. 
We set $K$ to 20, which means that 20 groups were randomly selected for study. 
Both $n$ and $m$, the number of masks used in separation and occlusion, were set to 4.

As shown in \figureautorefname \ref{fig:groups-boxplot}, the $\rho$-value of the Artcode group is much less dispersed than that of the non-Artcode groups, showing less distance between the median and mean $\rho$-value. 
The mean aggregated likelihood data ($\overline{\rho_G}$) (denoted by dashed lines in the boxes) shows $\overline{\rho_{G_a}}$ (0.136170) to be greater than all the $\overline{\rho_{G_{n_i}}}, i= 1\,\dots\,20$. 
This shows that, overall, the sum of probabilities of all image blocks of an Artcode image is greater than that of a non-Artcode image 
--- 
indicating that the MRs have not been violated. 
Because of the uncertain nature of supervised classification, the aggregated likelihood of an individual Artcode image is {\em not} always greater than that of a non-Artcode image 
--- 
the classifier may not predict inputs with 100\% accuracy. 
However, the statistical analysis of variations between the groups $G_a$ and $G_n$ provides evidence for the difference of the mean $\rho$-values between Artcode and non-Artcode groups, indicating no violation of the MRs.

\renewcommand{\arraystretch}{.7}
\begin{table}[t!]
\centering
\small\addtolength{\tabcolsep}{-1.5pt}
\caption{Results of verification statistical analyses.}
\label{tbl:verificationResults}
\begin{tabular}{l|P{2.5cm}|P{2.5cm}|P{3cm}|P{1.8cm}}
\hline
$\rho_{G_n}$ &
  One-way ANOVA &
  \begin{tabular}[c]{@{}c@{}}t-test (for \\ equal variances)\end{tabular} &
  \begin{tabular}[c]{@{}c@{}}ANOVA\_ranks\\ (Kruskal-Wallis test)\end{tabular} & 
  \begin{tabular}[c]{@{}c@{}}Dunnett's \\ test\end{tabular} \\ \hline
1      & \cellcolor{gray!75}0.020599 & \cellcolor{gray!75}0.020935 & \cellcolor{gray!75}0.026708  & 0.137527                    \\
2      & \cellcolor{gray!75}0.000929 & \cellcolor{gray!75}0.001055 & \cellcolor{gray!75}0.001055  & \cellcolor{gray!75}0.006173 \\
3      & \cellcolor{gray!75}0.020941 & \cellcolor{gray!75}0.021266 & \cellcolor{gray!75}0.027382  & \cellcolor{gray!75}0.005814 \\
4      & 0.151530  					 & 0.151567                    & \cellcolor{gray!75}0.014303  & \cellcolor{gray!75}0.022575 \\
5      & \cellcolor{gray!75}0.026392 & \cellcolor{gray!75}0.026708 & \cellcolor{gray!75}0.028257  & \cellcolor{gray!75}0.016151 \\
6      & \cellcolor{gray!40}0.082988 & \cellcolor{gray!40}0.082990 & \cellcolor{gray!75}0.011025  & \cellcolor{gray!75}0.025203 \\
7      & 0.563244                    & 0.563270                    & 0.252606                     & \cellcolor{gray!75}0.011663 \\
8      & \cellcolor{gray!75}0.004583 & \cellcolor{gray!75}0.004797 & \cellcolor{gray!75}0.004104  & \cellcolor{gray!75}0.023714 \\
9      & \cellcolor{gray!75}0.026213 & \cellcolor{gray!75}0.026354 & \cellcolor{gray!75}0.007432  & \cellcolor{gray!75}0.024604 \\
10      & 0.120316                    & 0.120447                    & \cellcolor{gray!40}0.088321 & \cellcolor{gray!75}0.004742 \\
11     & \cellcolor{gray!40}0.082720 & \cellcolor{gray!40}0.083057 & 0.119343                     & 0.198247                    \\
12     & \cellcolor{gray!75}0.006003 & \cellcolor{gray!75}0.006276 & \cellcolor{gray!75}0.008328  & \cellcolor{gray!75}0.013617 \\
13     & \cellcolor{gray!40}0.060633 & \cellcolor{gray!40}0.060637 & \cellcolor{gray!75}0.003846  & \cellcolor{gray!75}0.040912 \\
14     & \cellcolor{gray!40}0.059978 & \cellcolor{gray!40}0.060223 & \cellcolor{gray!40}0.051375  & \cellcolor{gray!75}0.029467 \\
15     & \cellcolor{gray!75}0.013469 & \cellcolor{gray!75}0.013771 & \cellcolor{gray!75}0.015104  & \cellcolor{gray!75}0.034990 \\
16     & \cellcolor{gray!75}0.007758 & \cellcolor{gray!75}0.008053 & \cellcolor{gray!75}0.011156  & \cellcolor{gray!40}0.089879 \\
17     & \cellcolor{gray!75}0.036771 & \cellcolor{gray!75}0.037150 & \cellcolor{gray!40}0.055180  & \cellcolor{gray!40}0.005291 \\
18     & 0.225829                    & 0.226021                    & 0.310495                     & \cellcolor{gray!75}0.042914 \\
19     & \cellcolor{gray!40}0.062216 & \cellcolor{gray!40}0.062425 & \cellcolor{gray!75}0.047335  & \cellcolor{gray!75}0.018983 \\
20     & 0.323313                    & 0.323352                    & \cellcolor{gray!40}0.085770  & \cellcolor{gray!75}0.029832 \\ \hline
median & 0.048375 & 0.048687 & 0.027045 & 0.024159 \\
mean   & 0.094821 & 0.095018 & 0.058456 & 0.039115 \\
min    & 0.000929 & 0.001055 & 0.001055 & 0.004742 \\
max    & 0.563244 & 0.563270 & 0.310495 & 0.198247 \\
std    & 0.137489 & 0.137422 & 0.083444 & 0.047734 \\ \hline
\end{tabular}
\end{table}

\tableautorefname\ref{tbl:verificationResults} presents the significance level ($\mathrm{p}$-values) of the difference between $\rho_{G_a}$ and $\rho_{G_{n_i}}$ under 
ANOVA, 
t-test (for unequal variances), 
ANOVA\_ranks, and 
Dunnett's test.
Descriptive statistics 
--- 
median, mean, minimum, maximum, and standard deviation (std)
---
for the $\mathrm{p}$-values are also included.
For ease of understanding, cells in the table are coloured to reflect the significance level:
$\mathrm{p} \le 0.05$ are shown in dark gray;
$0.05 < \mathrm{p} \le 0.10$ are in light gray; and
$\mathrm{p} > 0.10$ are in white. 
If the {\em null hypothesis} is defined as ``an MR is violated'', then small $\mathrm{p}$-values (typically below 0.05) indicate strong evidence against the null hypothesis
---
small $\mathrm{p}$-value indicate that neither of the two MRs have been violated.  
On the other hand, large $\mathrm{p}$-values indicate weak evidence to reject the null hypothesis: there is no significant difference between the mean $\rho$-values of the $\rho_{G_a}$ and $\rho_{G_n}$ groups, under the chosen significance level, suggesting that one or both of the MRs may have been violated and, thus, the RF-based original classifier may have defects.

As shown in \tableautorefname\ref{tbl:verificationResults}, the ANOVA $\mathrm{p}$-values range from 0.000929 to 0.563244, with a median of 0.048375.
Half of the $\rho_{G_n}$ groups show $\mathrm{p}$-values that are considerably less than 0.05, indicating that these groups ($\rho_{G_{n_i}}, i=1-3, 5, 8-9, 12,$ and $15-17$) are significantly different from $\rho_{G_a}$, under the significance level of 0.05 ($\alpha=0.05$). 
If we increase the alpha value to 0.1, then two thirds of non-Artcode groups $\rho_{G_{n_i}}$ have means that are significantly different from the Artcode group $\rho_{G_a}$. 
This result provides evidence that the difference between the two groups is not due to sampling errors or by chance. 
The $\mathrm{p}$-values of the remaining pairs are greater than 0.05, ranging from 0.059978 to 0.563244, indicating that there is no significant difference between the mean $\rho$-values of the two groups under $\alpha=0.05$.
This result can be explained by the diversity of the non-Artcode images in the Artcode dataset
---
some appear very similar to Artcode images, so-called ``Artcode-like'' images  \citep{liming2020thesis}. 
Therefore, the significance level of the difference between $\overline{\rho_{G_a}}$ and $\overline{\rho_{G_n}}$ may decrease if $\rho_{G_n}$ includes many Artcode-like images.
This will be discussed further in \sectionautorefname\ref{sec:discussion}.

The mean ANOVA $\mathrm{p}$-value is 0.094821, which is considerably larger than the median value of 0.048375. 
This indicates the skewness of the $\mathrm{p}$-values:
most $\mathrm{p}$-values approach the minimal $\mathrm{p}$-value, evidenced by the relatively higher standard deviation (0.137489). 
Although the mean $\mathrm{p}$-value is relatively high (greater than the commonly-used significance level of 0.05), the low median $\mathrm{p}$-value is evidence against the null hypothesis, reflecting the observed differences between $\overline{\rho_{G_a}}$ and most $\overline{\rho_{G_{n_i}}}$.

The one-way ANOVA results show that, even without assurance of equal variances and normality, $\rho_{G_{a}}$ is, to some extent, significantly different from $\rho_{G_{n_i}}$. 
Moreover, this significant difference was also observed under the assumptions of unequal variances and non-normality. 
The t-test (for unequal variances) has almost equivalent results to ANOVA (with only a negligible increase in $\mathrm{p}$-values), thus supporting the same conclusion as ANOVA.

\tableautorefname \ref{tbl:verificationResults} also reports the results of the Kruskal-Wallis tests (ANOVA\_ranks), which are suitable for non-normally distributed data. 
The ANOVA\_ranks $\mathrm{p}$-values are generally lower than those of ANOVA, ranging from 0.001055 to 0.310495, with a median of 0.027045 (which is less than the commonly-used $\alpha$-value of 0.05). 
13 groups (1-6, 8-9, 12-13, 15-16, and 19) have $\mathrm{p}$-values below 0.05. 
Compared with the ANOVA and t-test (for unequal variances) results, ANOVA\_ranks has a considerably lower mean $\mathrm{p}$-value (0.058456), which is only slightly greater than the $\alpha$-value of 0.05. 
The dispersion of $\mathrm{p}$-values is also lower, with a smaller standard deviation of 0.083444.
The ANOVA\_ranks results confirm the significant differences between the means of $\rho_{G_a}$ and $\rho_{G_{n_i}}$ under the assumption of non-normality. 
This phenomenon could be explained by the ranked data type of the $\rho$-values: 
the $\rho$-values are not completely continuous, or normally distributed, but somehow show ``ranks'' in the proposed MR-augmented framework.

The $\mathrm{p}$-values for one-way ANOVA, ANOVA\_ranks, and  
t-test (for unequal variances) were calculated in separate comparisons.
To alleviate the influence of this setting, and to consolidate the conclusion, we also conducted a multiple comparison test, Dunnett's test, to compare the Artcode group $\rho_{G_a}$ and the 20 non-Artcode groups $\rho_{G_{n_i}}$. 
Because the experiment studied the difference between $\rho_{G_a}$ and $\rho_{G_{n}}$, only the $\mathrm{p}$-values for comparisons between $\rho_{G_a}$ and each $\rho_{G_{n_i}}$ are presented in \tableautorefname \ref{tbl:verificationResults}.  
As can be seen from the table, Dunnett's test provides more evidence for significant differences between $\rho_{G_a}$ and $\rho_{G_n}$, with the $\mathrm{p}$-values ranging from 0.004742 to 0.198247, and a median of 0.024159. 
16 of the 20 groups were significantly different ($\alpha = 0.05$) from the Artcode group.
In terms of mean and standard deviation, Dunnett's test had the lowest mean (0.039115) and standard deviation (0.047734) among all four tests. 
The results of the Dunnett's test thus confirm the significant difference between the Artcode group and non-Artcode groups.

Although none of the four test methods produced 20 $\mathrm{p}$-values below 0.05, overall, the results in \tableautorefname \ref{tbl:verificationResults} show significant differences between the mean aggregated likelihoods of the Artcode and non-Artcode groups. 
Considering the uncertain nature of the predictor (the classifier) and the innate {\em variance} of random forests, the experimental results indicate no reason to consider the implementation faulty
---
the results indicate that neither MR has been violated.
The next section will present the second study to evaluate the performance of the MR-augmented classifier, showing the enhanced performance of MR-augmented classifiers over non-augmented classifiers.

\subsection{Study 2 -- Enhancement}\label{subsec:enhancement}
\subsubsection{Experimental setting}\label{subsec:experimentalSetting}
According to the framework in \figureautorefname\ref{fig:augmentedClassifierFramework}, we used Matlab to implement MR-augmented versions of classifiers that use random forests and support vector machines.
The RF-based MR-augmented, SVM-based MR-augmented, RF-based non-MR-augmented (original) and SVM-based non-MR-augmented classifiers are denoted Aug-RF, Aug-SVM, Ori-RF and Ori-SVM, respectively.
Cross-validation techniques were used to evaluate and compare the performance of these classifiers, with the Artcode dataset being used as the evaluation dataset.

Because random forests and SVM are used for the classification algorithms, the performance naturally has a certain level of variation in each execution 
---
due to RF's random variable selection from the feature vector, and SVM's sub-optimisation because of the limited number of computational iterations.
Multiple runs of cross-validation were therefore conducted to obtain the average performance. 
Because the dataset was imbalanced, with more non-Artcode than Artcode samples, we needed an appropriate group of measurements that could effectively deal with evaluation using imbalanced datasets to provide an informative view of the performance of the MR-augmented classifiers:
Precision, recall, accuracy, the TNR (true negative rate), the $F_\beta$ measure, and the MCC (Matthews Correlation Coefficient) \citep{matthews1975comparison} were all employed as evaluation metrics. 

Precision is a measure of the correctness of those classified as Artcodes, whereas recall is a measure of completeness (how many of the true Artcodes were correctly classified). 
These two measures focus on positive examples and predictions, and their importance varies from one learning task to another. 
With Artcode classification, recall is more important than precision because recognising the presence of all Artcodes in the scene is a prerequisite to the follow-up decoding that triggers the digital information.

TNR measures how many non-Artcode samples are correctly classified. Accuracy, F$_\beta$, and MCC measure the overall performance of the classifier. Accuracy is the overall proportion of correct predictions, for both the positives (Artcodes) and negatives (non-Artcodes). However, accuracy is sensitive to size differences among classes, and, in our study, may have been influenced by the imbalanced class sizes. The F$_2$ measure is a special instance of the F$_\beta$ measure with $\beta = 2$, where $\beta$ is a value allocating $\beta$ times as much importance to recall as to precision.
F$_2$ uses a weighted average of precision and recall to evaluate the classification effectiveness, giving twice ($\beta=2$) as much importance to recall as to precision. In contrast to accuracy, the F$_2$ measure and MCC provide more insight into the performance of a classifier. However, compared with MCC, F$_2$ can be sensitive to data distribution. 
MCC is, in essence, a correlation coefficient between the observed and predicted classifications, incorporating true and false positives and negatives. It remains effective even if the dataset is imbalanced, and is generally regarded as one of the best measures for classification performance evaluation \citep{powers2011evaluation}.

Two thresholds, $t_1$ and $t_2$, were studied in the experiment, as was their impact on the augmented classifiers. 
The given values in the weight vector $\vec{w}$ affect the selection of the values of $t_1$ and $t_2$. 
According to \equationautorefname \ref{eq:occlusion2}, the weights of image blocks generated by occlusion are greater than those generated by separation. 
In this experiment, four masks were used for both separation and occlusion ($n = m = 4$), resulting in both the prediction vector $\vec{p}$ and the weight vector $\vec{w}$ being 8-dimensional. 
Based on empirical examinations of assigning different values to $\vec{w}$, 
we assigned a value of $0.1$ to both $w_{s_a}$ and $w_{s_n}$, and 
a value of $0.15$ to both $w_{o_a}$ and $w_{o_n}$. 
In order to achieve quantisation and computational convenience of the value of aggregated likelihood $\rho$, the numbers $1$ and $0$ were used in the prediction vector $\vec{p}$ to represent the Artcode and non-Artcode classes, respectively.

\subsubsection{Results}\label{subsec:results}
\begin{figure}[t!]
        \centering
         \begin{subfigure}[b]{0.48\textwidth}
                \includegraphics[width=\textwidth]{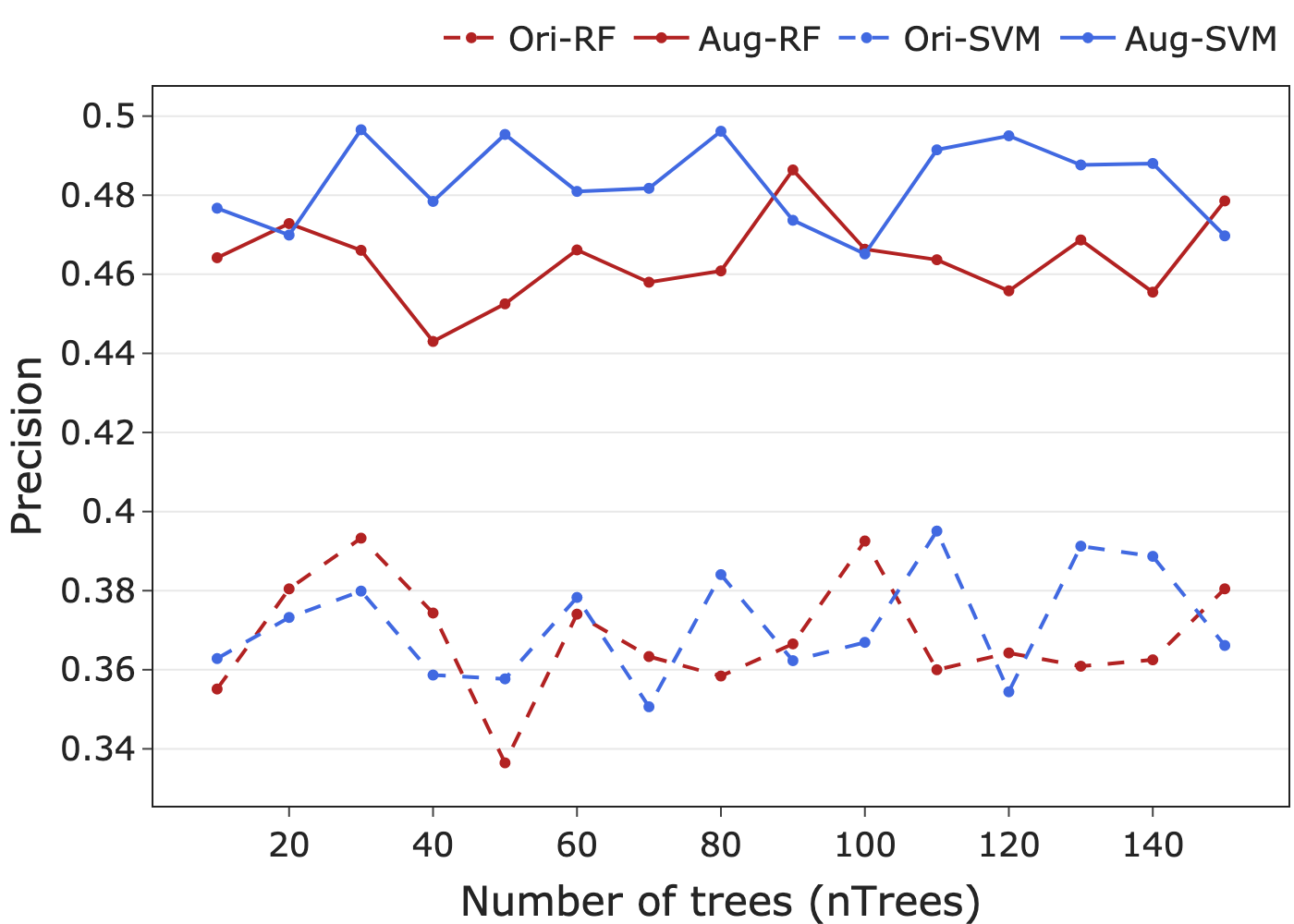}
                \caption{Precision}
                \label{fig:evaluationMetrics_1_a}
        \end{subfigure}  
        \begin{subfigure}[b]{0.48\textwidth}
                \includegraphics[width=\textwidth]{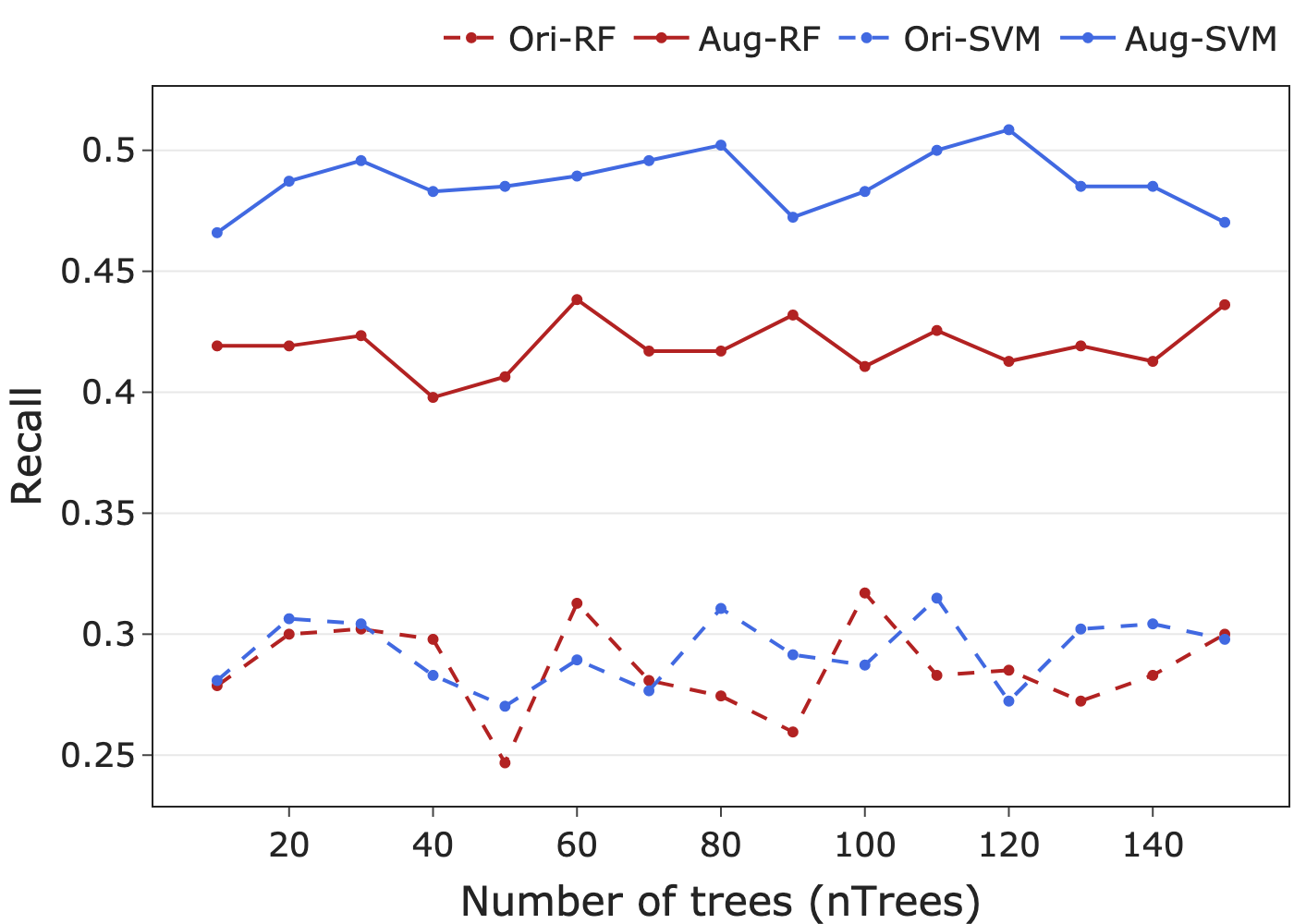}
                 \caption{Recall}
                 \label{fig:evaluationMetrics_1_b}
        \end{subfigure}  
        \begin{subfigure}[b]{0.48\textwidth}
                \includegraphics[width=\textwidth]{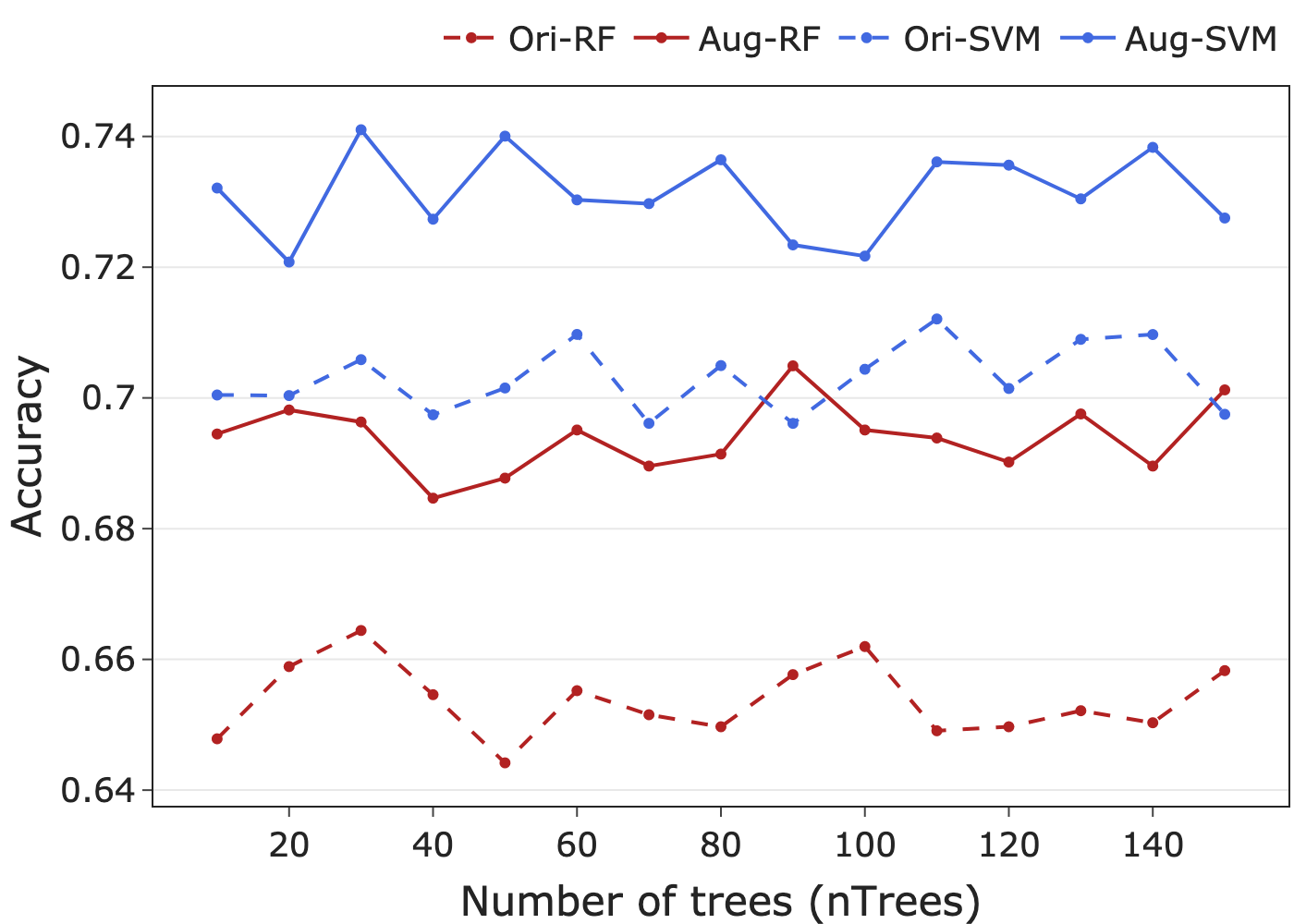}
                 \caption{Accuracy}
                 \label{fig:evaluationMetrics_1_c}
        \end{subfigure} 
        \begin{subfigure}[b]{0.48\textwidth}
                \includegraphics[width=\textwidth]{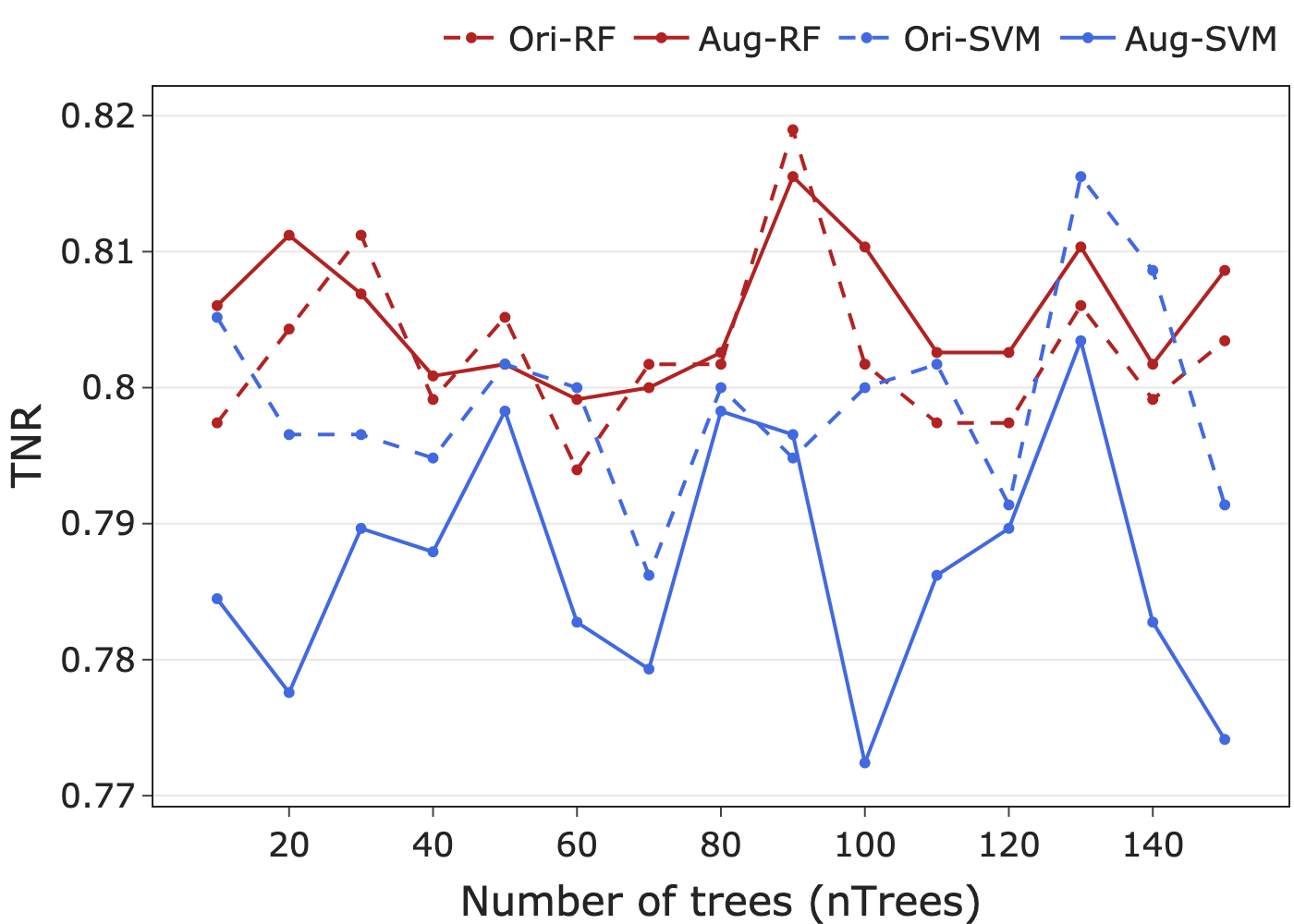}
                \caption{True negative rate (TNR)}
                 \label{fig:evaluationMetrics_1_d}
        \end{subfigure} 
        \begin{subfigure}[b]{0.48\textwidth}
                \includegraphics[width=\textwidth]{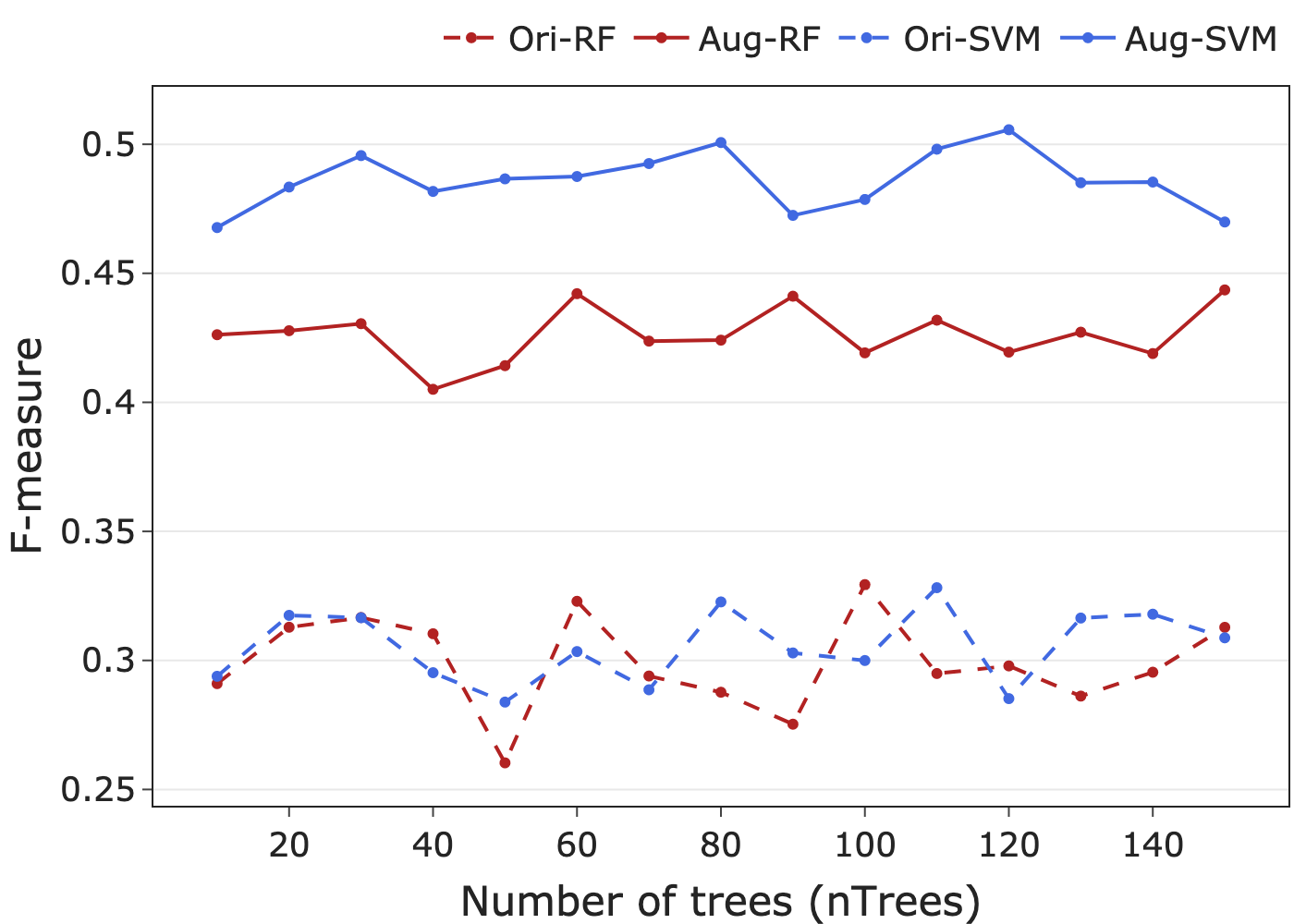}
                \caption{F$_2$ measure}
                 \label{fig:evaluationMetrics_1_e}
        \end{subfigure}  
        \begin{subfigure}[b]{0.48\textwidth}
                \includegraphics[width=\textwidth]{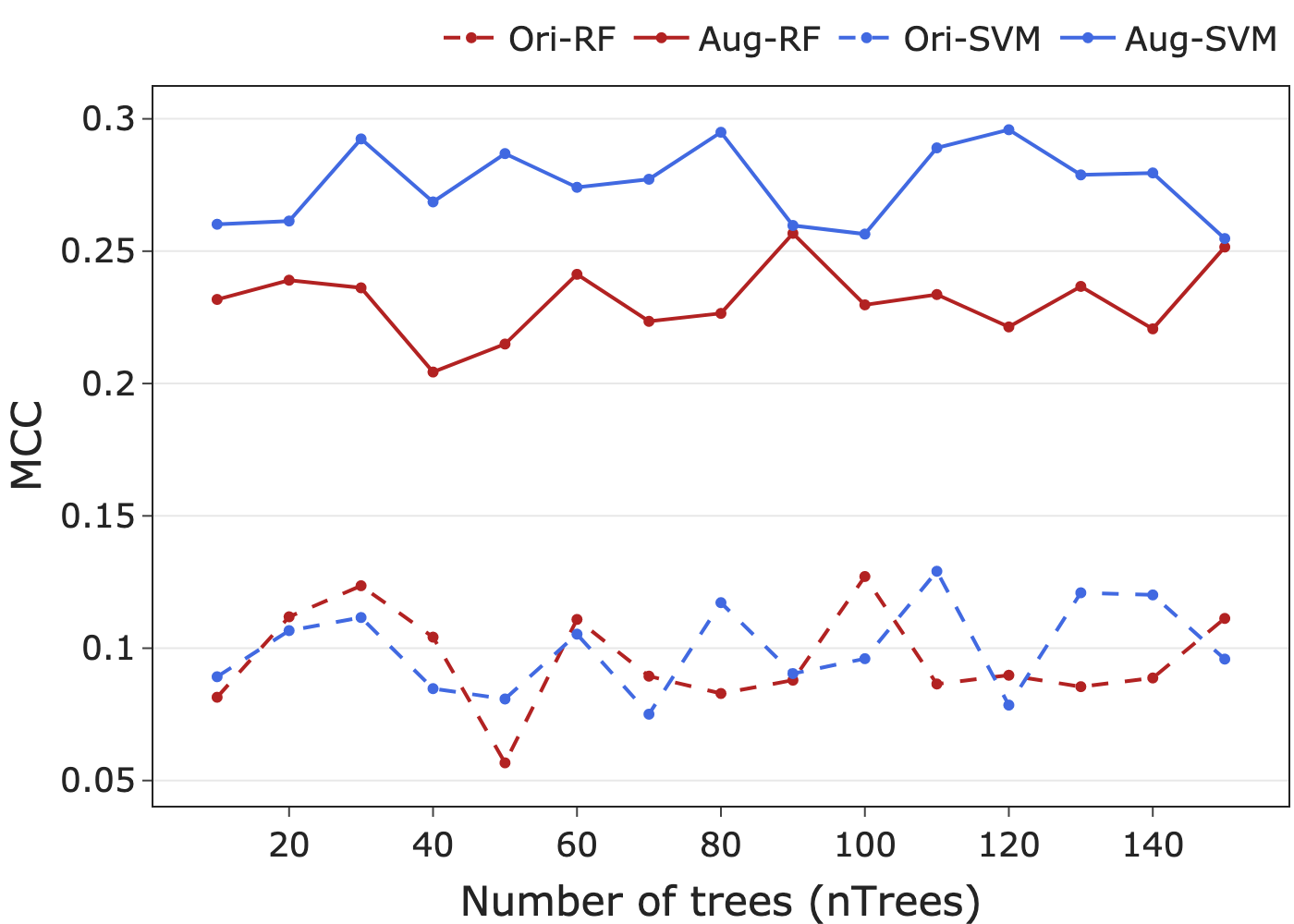}
                \caption{Matthews correlation coefficient (MCC)}
                 \label{fig:evaluationMetrics_1_f}
        \end{subfigure}   
        \caption{Performance comparison between RF and SVM-based classifiers with different values of $nTrees$ and $t_1, t_2=0.2$.}
        \label{fig:evaluationMetrics_1}
\end{figure}

\begin{figure}[t!]
        \centering 
         \begin{subfigure}[b]{0.48\textwidth}
                \includegraphics[width=\textwidth]{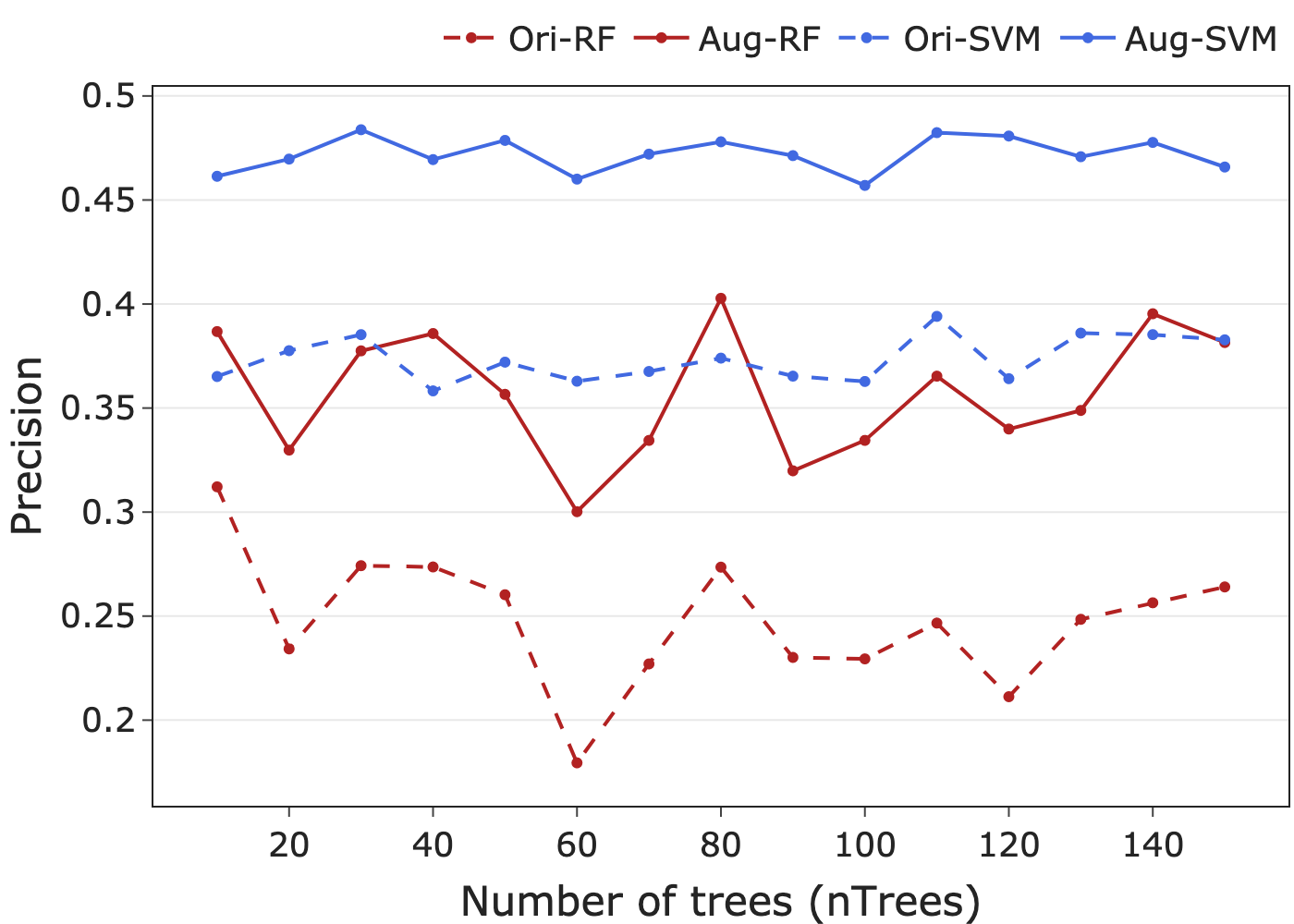}
                \caption{Precision}
                \label{fig:evaluationMetrics_2_a}
        \end{subfigure}  
        \begin{subfigure}[b]{0.48\textwidth}
                \includegraphics[width=\textwidth]{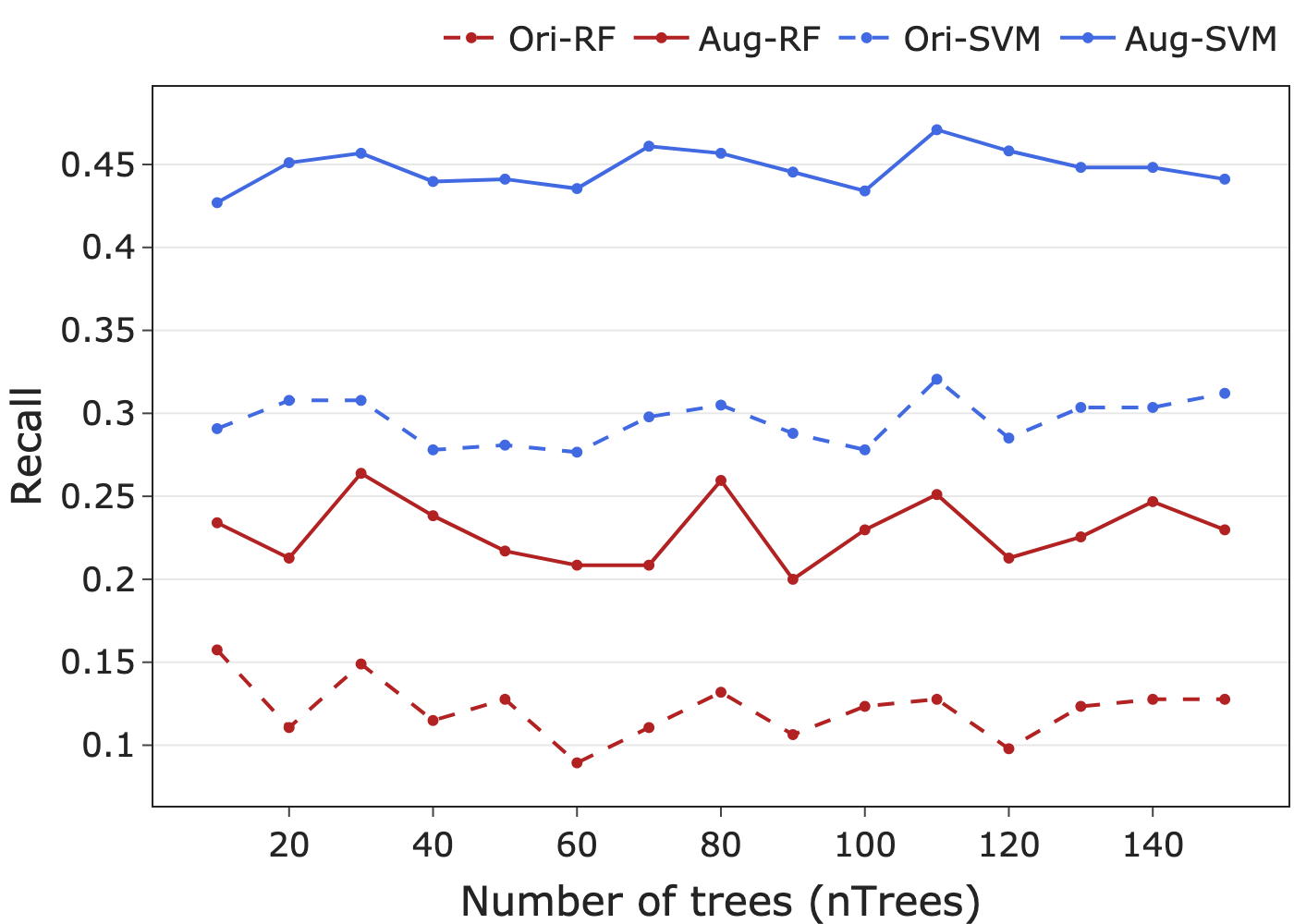}
                \caption{Recall}
                 \label{fig:evaluationMetrics_2_b}
        \end{subfigure}  
        \begin{subfigure}[b]{0.48\textwidth}
                \includegraphics[width=\textwidth]{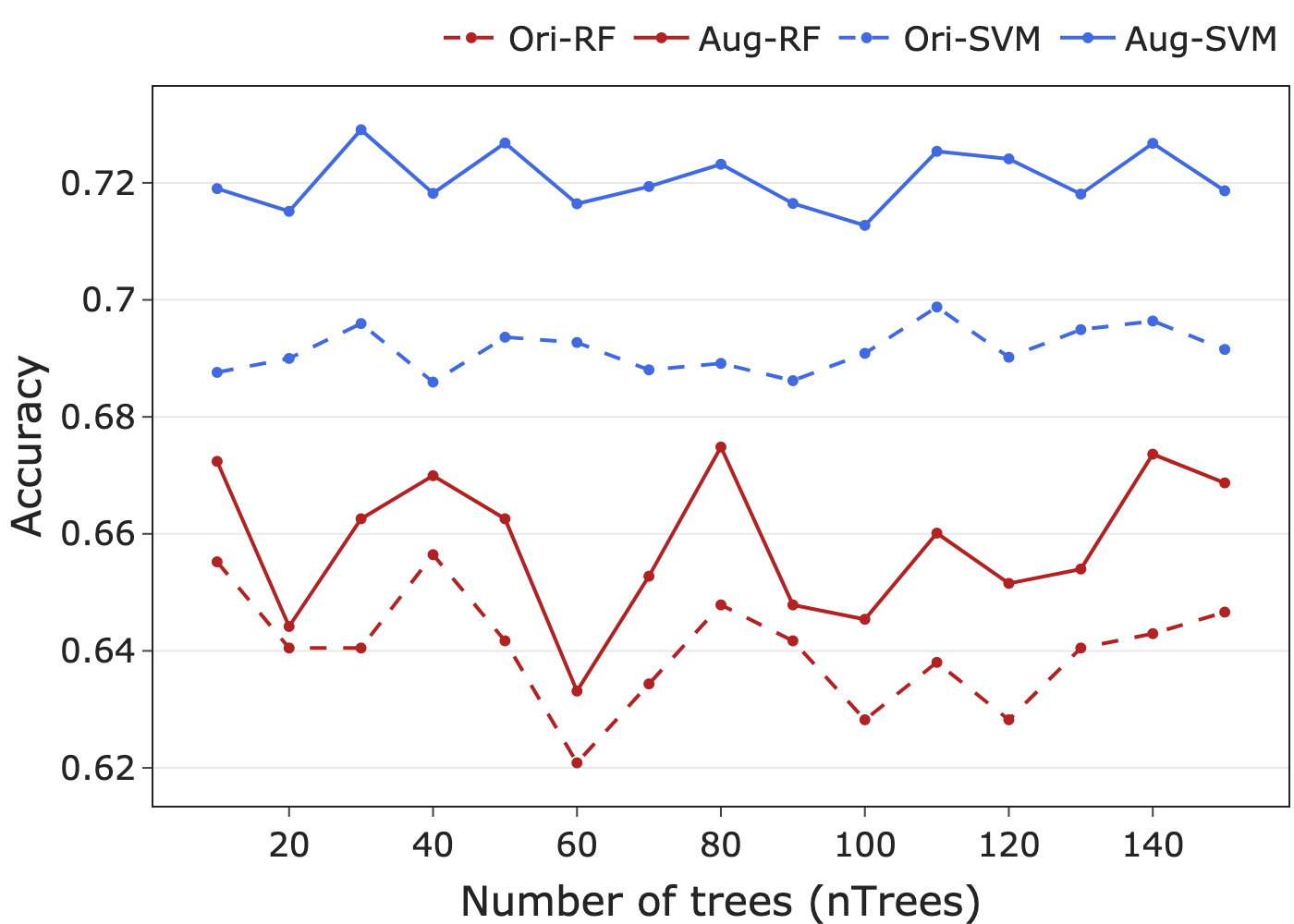}
                \caption{Accuracy}
                 \label{fig:evaluationMetrics_2_c}
        \end{subfigure}
        \begin{subfigure}[b]{0.48\textwidth}
                \includegraphics[width=\textwidth]{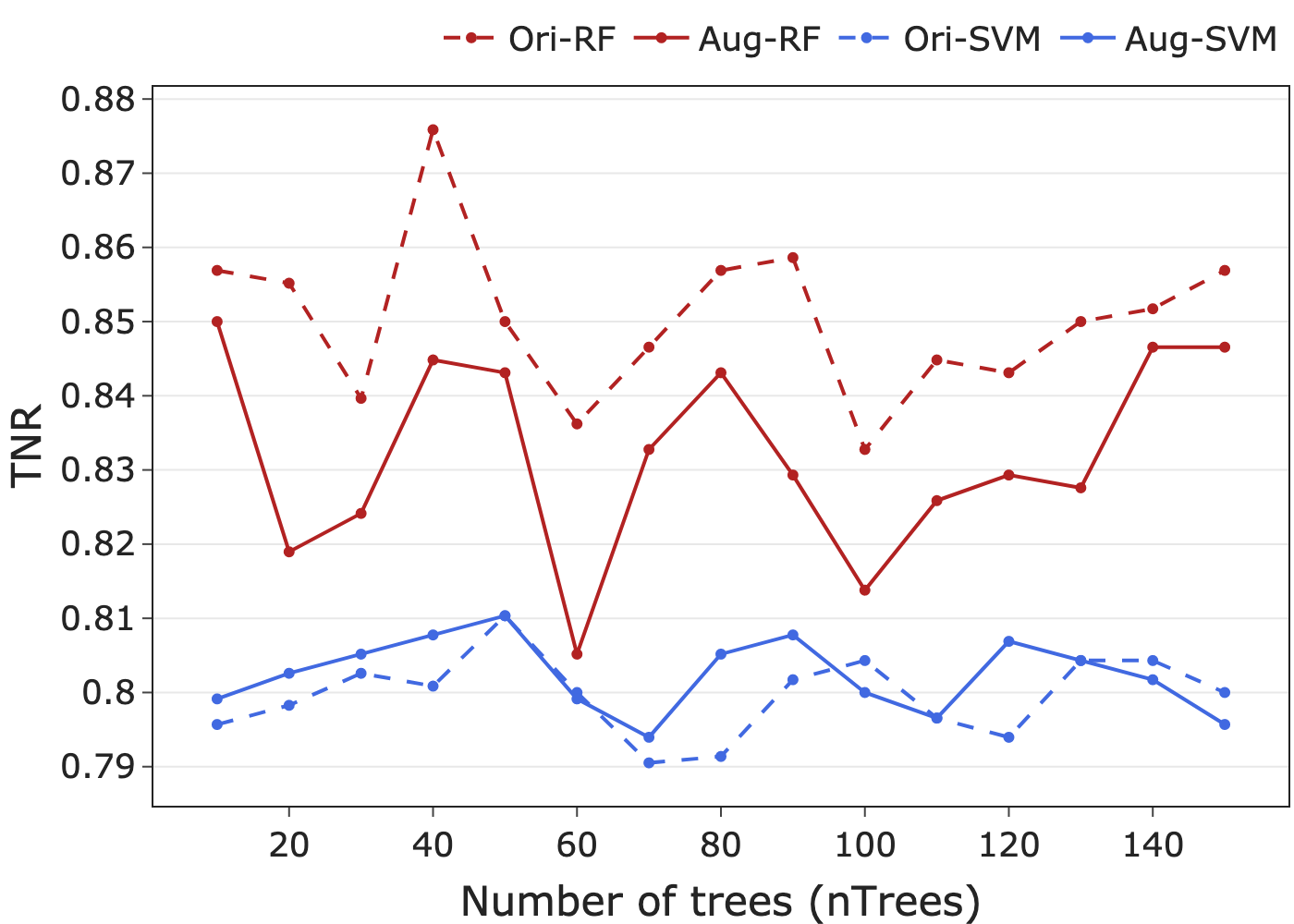}
                \caption{True negative rate (TNR)}
                 \label{fig:evaluationMetrics_2_d}
        \end{subfigure}
        \begin{subfigure}[b]{0.48\textwidth}
                \includegraphics[width=\textwidth]{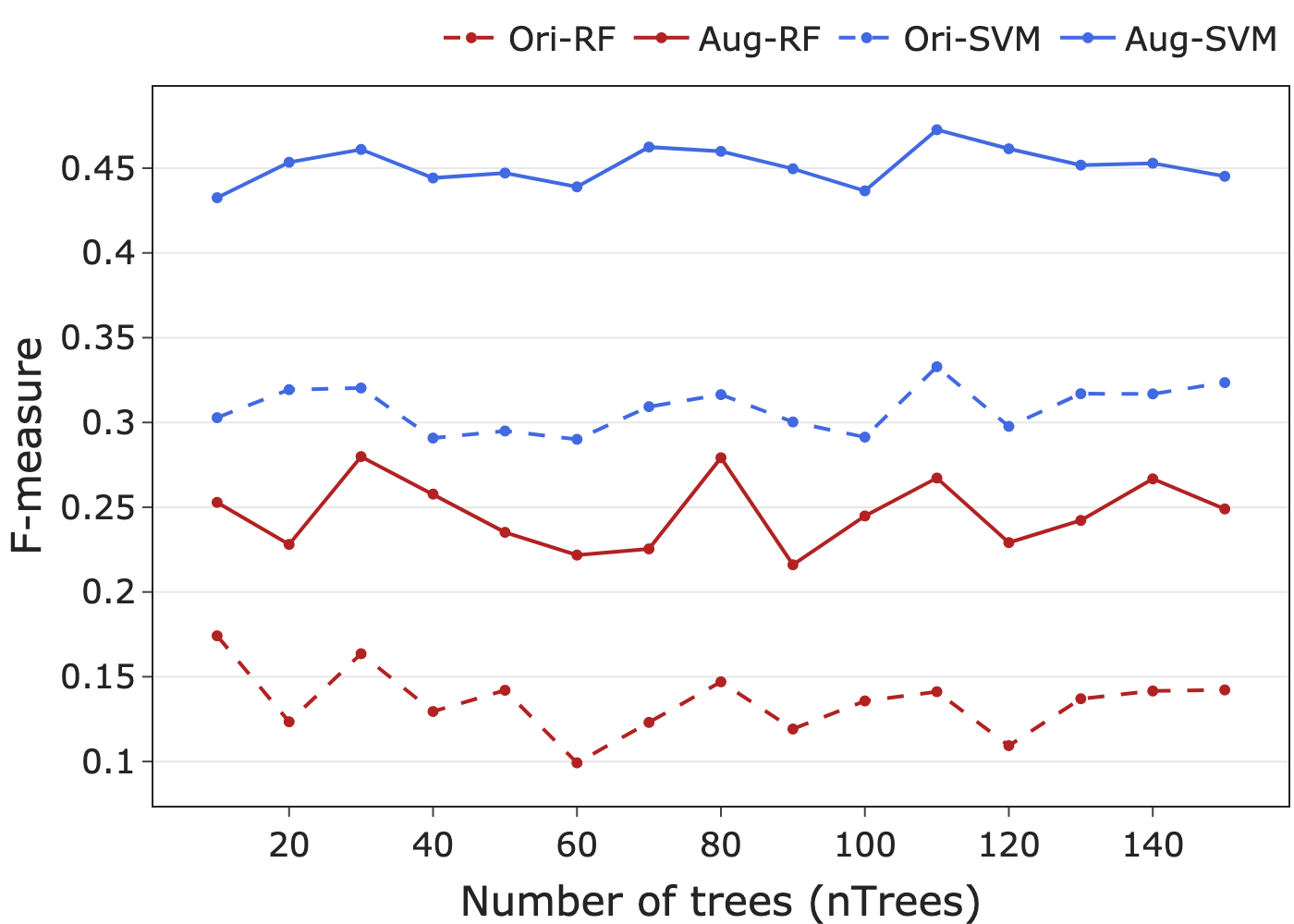}
                \caption{F$_2$ measure}
                 \label{fig:evaluationMetrics_2_e}
        \end{subfigure}   
        \begin{subfigure}[b]{0.48\textwidth}
                \includegraphics[width=\textwidth]{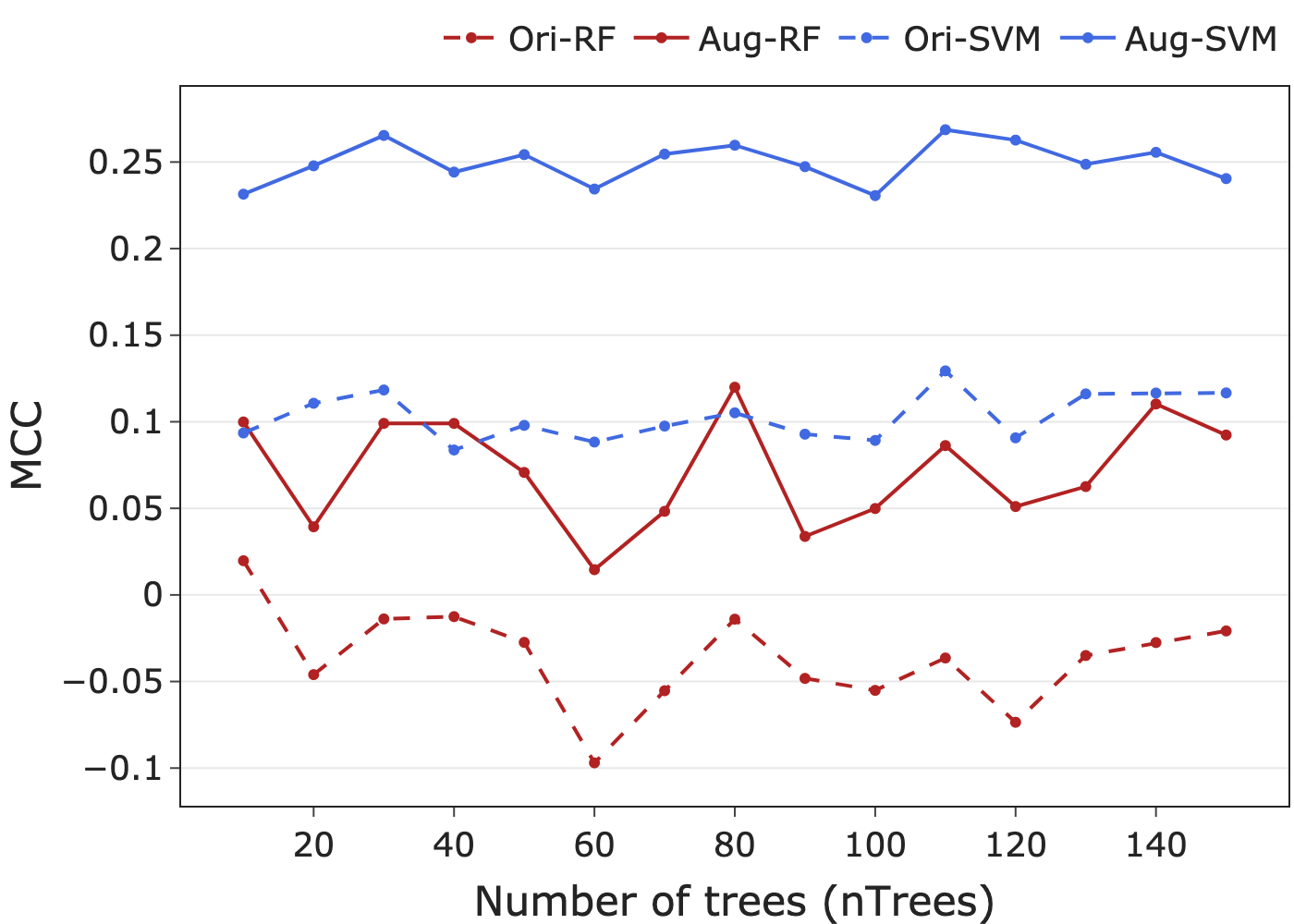}
                \caption{Matthews correlation coefficient (MCC)}
                 \label{fig:evaluationMetrics_2_f}
        \end{subfigure}   
        \caption{Performance comparison between the RF- and SVM-based classifiers with different $nTrees$ values and $t_1=0.15, t_2=0.3$.}
         \label{fig:evaluationMetrics_2}
\end{figure}

All performance metric values reported are the average values calculated from five executions of $k$-fold cross-validation. 
Two combinations of the two thresholds $t_1$ and $t_2$, in conjunction with different numbers of decision trees, were used to study the impact of the classifiers' tuning parameters. 
Because $nTrees$ (the number of decision trees used in the RF-based classifiers) is not a tuning parameter of the SVM classifiers, for the sake of comparison, the SVM classifier values for each $nTrees$ value are only the average of five runs of $k$-fold cross-validation. 
Higher values in \figuresautorefname \ref{fig:evaluationMetrics_1} and \ref{fig:evaluationMetrics_2} indicate better performance. 
\figuresautorefname\ref{fig:evaluationMetrics_1} and \ref{fig:evaluationMetrics_2} show a consistent performance across different values of $nTrees$ for all six evaluation metrics: 
This means that the Aug-RF classifier (unbroken red) is not sensitive to changes in the value of $nTrees$, a characteristic inherited from the original RF classifier (dashed red).

\paragraph{MR-augmented {\em versus} non-MR-augmented classifiers}\label{par:compAugmentedClassifiers}
We studied the performance difference between the augmented (Aug-) and original (Ori-) classifiers, and also compared the performance of the classifiers based on random forests (-RF) with that of those based on support vector machines (-SVM).

Using various values of $nTrees$ and fixed values of the thresholds $t_1$ and $t_2$, the MR-augmented classifiers (Aug-RF and Aug-SVM) outperformed the original classifiers (Ori-RF and Ori-SVM) in terms of precision, recall, accuracy, F$_2$, and MCC. 
They also outperformed the original classifiers in terms of recall, precision, and F$_2$ measure for threshold combinations of $t_1=t_2=0.2$, and $t_1=0.15$, $t_2=0.3$, showing improved predictive performance in classification of the positive class (Artcodes). 
This improvement is important because Artcode classification requires higher accuracy when predicting Artcodes.

When predicting the negative class (non-Artcodes), as measured by TNR, the MR-augmented classifiers appear slightly influenced by different values of the thresholds ($t_1$ and $t_2$), which can be seen in the slight difference in TNR values for the original and augmented classifier in \figuresautorefname\ref{fig:evaluationMetrics_1}\subref{fig:evaluationMetrics_1_d} and \ref{fig:evaluationMetrics_2}\subref{fig:evaluationMetrics_2_d}: 
for $t_1=t_2=0.2$, the augmented classifier TNR values are similar to those for the original; 
but for $t_1=0.15$, $t_2=0.3$, they are less effective. 
This is different to the other evaluation metrics, which all show that the augmented classifiers outperform the original ones for both threshold combinations. 
A reason for this, partly as described in Section \ref{subsec:rectification}, is that when $t_1$ equals $t_2$, the augmented classifier does not directly use the prediction result of the original classifier. 
Another reason is the careful selection of threshold $t_1$: 
lower values of $t_1$ mean that the augmented classifier predicts the input image depending on the MRs only when they can adjust prediction with a relatively high confidence
---
otherwise, the augmented classifier uses the original prediction result. 
Thus, thresholds $t_1$ and $t_2$ can be used as tuning parameters for the performance of the MR-augmented classifier for both the positive and negative class.

Accuracy and MCC assess the overall performance of the classifier.
As shown in \figuresautorefname\ref{fig:evaluationMetrics_1}\subref{fig:evaluationMetrics_1_c} and \ref{fig:evaluationMetrics_2}\subref{fig:evaluationMetrics_2_c}, for both threshold combinations, the augmented classifiers have slightly better Accuracy than the original classifier, with an average increase of approximately 2-3\%. 
Although the MR-augmented classifiers show improved performance in the Artcode class, the small percentage of Artcodes in the dataset does not contribute strongly to the overall accuracy in evaluation, which is determined by both true positives and true negatives. 
In contrast, MCC is a more informative measure of overall performance, even when the dataset is imbalanced. 
As shown in \figuresautorefname\ref{fig:evaluationMetrics_1}\subref{fig:evaluationMetrics_1_f} and \ref{fig:evaluationMetrics_2}\subref{fig:evaluationMetrics_2_f}, the augmented classifiers obtain about a 10-20\% increase over the original classifiers. 
This improvement is much more noticeable when comparing Aug-SVM with the Ori-SVM classifier, showing an overall improved performance of the MR-augmented classifier. 
However, the values of F$_2$ and MCC for all classifiers are relatively low. 
This is due to the imbalance of the dataset used in the evaluation, with a much greater number of negative examples than positive ones.

Both the original and MR-augmented classifiers achieve high true negatives (TN), approximately 0.82--0.85, as presented in \figuresautorefname\ref{fig:evaluationMetrics_1}\subref{fig:evaluationMetrics_1_d} and \ref{fig:evaluationMetrics_2}\subref{fig:evaluationMetrics_2_d}.
However, they also have very low true positives (TP), approximately 0.3--0.4, which can be observed from the low precision 
(\figuresautorefname\ref{fig:evaluationMetrics_1}\subref{fig:evaluationMetrics_1_a} and \ref{fig:evaluationMetrics_2}\subref{fig:evaluationMetrics_2_a}) and 
recall 
(\figuresautorefname\ref{fig:evaluationMetrics_1}\subref{fig:evaluationMetrics_1_b} and \ref{fig:evaluationMetrics_2}\subref{fig:evaluationMetrics_2_b}) 
results. 
If $TN = 0.85$ and $TP = 0.3$, then $FN = (1 - TN) = 0.15$ and $FP = (1 - TP) = 0.7$, and MCC can be calculated as: 
\begin{equation}\label{eq:mcc}
    MCC = \frac{TP \times TN - FP \times FN}{\sqrt{(TP + FP)(TP + FN)(TN + FP)(TN + FN)}}
\end{equation}
The MCC is a very low value, 0.1796. 
This illustrates how MCC is an effective measurement for evaluating the performance of a classifier on an imbalanced dataset.

As can be seen from \figuresautorefname\ref{fig:evaluationMetrics_1} and \ref{fig:evaluationMetrics_2}, the precision and recall values of all classifiers are relatively low, and the TNR values are comparatively high. 
This is due to the imbalance in the Artcode dataset, which includes many more negatives. 
On the one hand, more weight is given to the non-Artcode class by feeding more information to the classification model in the training stage, resulting in a classifier with low recall evaluation (good non-Artcode classification, but poorer Artcode classification). 
On the other hand, the small percentage of Artcodes in the dataset results in the low precision evaluation of both classifiers. 
Conversely, the large proportion of non-Artcode images in the dataset (and the good non-Artcode prediction of the classifier) lead to relatively high TNR values, as shown in \figuresautorefname\ref{fig:evaluationMetrics_1}\subref{fig:evaluationMetrics_1_d} and \ref{fig:evaluationMetrics_2}\subref{fig:evaluationMetrics_2_d}.

\paragraph{RF-based {\em versus} SVM-based classifiers}
The SVM-based classifiers (blue lines) achieve better performance than the RF-based classifiers (red lines), as shown in \figuresautorefname\ref{fig:evaluationMetrics_1} and \ref{fig:evaluationMetrics_2}, with an approximately 5-10\% increase in terms of almost all performance evaluation measurements (not for TNR). 
The tradeoff between the precision and TNR of the SVM-based classifiers can be adjusted by the misclassification matrix \citep{cortes1995support} employed in SVM. 
Considering the greater importance of recall than precision in this application, this experiment assigned higher values to the cost of classifying an Artcode as a non-Artcode, resulting in a classifier that enables better Artcode prediction.

The better performance of the SVM-based classifiers is also evidenced by the higher values of the Aug-SVM classifier than the Aug-RF classifier. 
However, when the classifiers use the same classification method (SVM or RF), the MR-augmented version outperformed the original (non-augmented) version of the corresponding classifier. 
This indicates that the introduction of MRs into supervised classification models actually improves the performance of the original classifiers, regardless of whether SVM or RF is used.

Overall, the Aug-SVM classifier obtained the best performance, especially when considering that SVM runs much faster than the random forests classifier. 
The MR-augmented classifiers outperformed the original classifiers in terms of all the evaluation measures. 
This improved performance is sensitive to the values of the thresholds $t_1$ and $t_2$, but not to the value of $nTrees$, or the choice of classification method. 
As discussed in \sectionautorefname \ref{par:compAugmentedClassifiers}, thresholds $t_1$ and $t_2$ influence the performance of the augmented classifier, with different combinations determining the impact the MRs have on adjusting the original classification. 
Careful selection of the values of the tuning parameters --- 
the thresholds $t_1$ and $t_2$ --- 
is therefore vital to fine-tune the results of the original classifier and obtain the enhanced performance.

\section{Analysis and discussion}\label{sec:discussion}
\begin{table}[ht]
	\caption{Results of rectification analysis.}
	\label{tbl:rectificationComparison}
	\centering
	\setlength{\tabcolsep}{1.8pt}
	\footnotesize{
	\begin{tabular}{lllllll}
	\hline
	          &   & \multicolumn{2}{c}{Aug-RF Rectifications} & ~~~ &\multicolumn{2}{c}{Aug-SVM Rectifications} \\  \cmidrule{3-4}\cmidrule{6-7}  
	     Class & Amount~~~ & Correct & Incorrect & & Correct & Incorrect  \\ 
         \hline
 		Artcode & 47    & 13.3 (28.3\%)     & 1.9 (4.04\%)   &   & 13.8 (29.36\%)    & 4 (8.51\%) \\
 		non-Artcode & 116  & 7.3 (6.3\%)    & 15.6 (13.45\%)  & & 6.4 (5.52\%)   & 9.2 (7.93\%)\\ 
	\hline
	\end{tabular}}
	\centering
\end{table}

\subsection{Analysis of the rectification stage}
\begin{figure}[t]
	\centering
	\includegraphics[width=\textwidth]{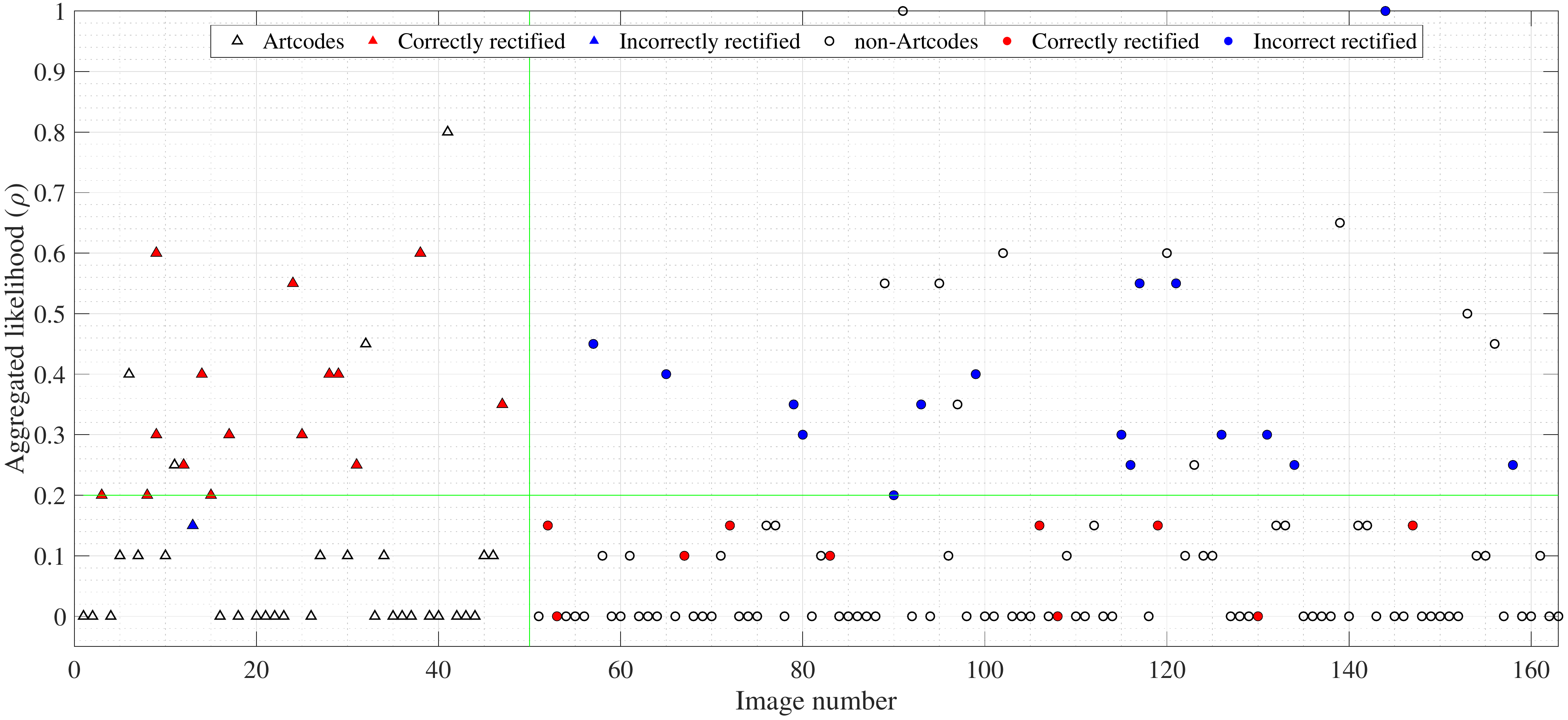}
	\caption{Illustration of rectification distribution of the RF-based MR-augmented classifier. 
	The graph is generated from one round of cross-validation of the RF-based MR-augmented classifier with $nTrees = 30$, and $t_1 = t_2 =  0.2$. 
	This graph is split into left and right areas separated by a green vertical line, where the left and right area are an illustration of the aggregated likelihood  ($\rho$-value) of Artcode ($\circ$) and non-Artcode ({\tiny $\triangle$}) images. 
	The horizontal green line is the predefined thresholds ($t_1$ and $t_2$):
	It separates the graph into upper and lower zones. The samples in the upper zone ($\ge t_2$ ) are rectified as {\em Artcodes}, whereas the samples in the lower zone ($\le t_1$) are labelled as {\em non-Artcodes} in the rectification stage of the MR-augmented classifier.	
	Therefore, the two tuning parameters, thresholds $t_1$ and $t_2$, of the MR-augmented classifier control whether or not to rectify more ``Artcode-like'' samples. 
	Correctly and incorrectly rectified predictions are highlighted in red and blue, respectively.}
    \label{fig:rectificationIllustration}
\end{figure}

In order to reveal how the fine-tuning (rectification layer) stage operates, and how the improved performance is achieved, we performed ten rounds of cross-validation runs using both the RF-based and SVM-based MR-augmented classifiers on all samples in the Artcode dataset. 
\tableautorefname \ref{tbl:rectificationComparison} shows the average correct and incorrect rectifications by the MR-augmented classifiers over these ten executions of 5-fold cross-validation. 
\figureautorefname \ref{fig:rectificationIllustration} shows $\rho$-values of all Artcodes ($\triangle$) and non-Artcodes ($\circ$) of one execution of cross-validation, where correct and incorrect rectifications are highlighted in red and blue, respectively. 
As illustrated in \figureautorefname \ref{fig:rectificationIllustration} and \tableautorefname \ref{tbl:rectificationComparison}, the two MR-augmented classifiers correctly rectified an average of 28.3\% and 29.36\% of the Artcode predictions, but incorrectly adjusted an average of 4.04\% and 8.51\% of the Artcodes to non-Artcodes. 
This higher correct rectification percentage contributed to the higher true positive rate 
--- 
a key factor in the evaluation of a classifier in terms of recall and precision. 
However, the classifiers performed slightly worse on the non-Artcode class: 
the RF-based MR-augmented classifier had an average of 6.63\% correct and 13.45\% incorrect rectifications, and the SVM-based MR-augmented classifier obtained an average of 5.52\% correct and 7.93\% incorrect rectifications. 
This explains why the MR-augmented classifiers have a relatively lower true negative rate (TNR), as shown in \figuresautorefname\ref{fig:evaluationMetrics_1}\subref{fig:evaluationMetrics_1_d} and \ref{fig:evaluationMetrics_2}\subref{fig:evaluationMetrics_2_d}, but higher precision (\figuresautorefname\ref{fig:evaluationMetrics_1}\subref{fig:evaluationMetrics_1_a} and \ref{fig:evaluationMetrics_2}\subref{fig:evaluationMetrics_2_a}) and recall (\figuresautorefname\ref{fig:evaluationMetrics_1}\subref{fig:evaluationMetrics_1_b} and \ref{fig:evaluationMetrics_2}\subref{fig:evaluationMetrics_2_b}). 
Overall, the average correct rectification percentage is 1.91\% ($\frac{13.3-1.9+7.3-15.6}{47+116} = 1.91\%$) for the Aug-RF classifier and 4.29\% ($\frac{13.8 - 8.51 + 6.4 - 9.2}{47 + 116} = 4.29\%$) for the Aug-SVM classifier, indicating that 1.91\% and 4.29\% of incorrect predictions by the RF-based and SVM-based original classifier were corrected by their respective MR-augmented classifiers. This explains how the improved performance of the MR-augmented classifiers was obtained: the rectification stage can rectify misclassifications (mainly false negatives) made by the original classifiers, albeit at the expense of comparatively fewer incorrect rectifications of true negatives.

The superior rectification performance of the MR-augmented classifiers on the Artcode examples shows that Artcode blocks are more likely to preserve the topological structure than non-Artcode blocks. 
Therefore, although the two MRs may violate the properties of Artcode images, the aggregated predictions of image blocks of Artcodes are more informative than the predictions of the entire image. 
This property may not be preserved for non-Artcodes, which have no predefined topological characteristics. 
The MR-augmented classifier adjusts prediction results in the rectification stage only if the new evidence collected is strong enough to accept, which is determined by comparing the aggregated likelihood of the predictions of image blocks with the given thresholds $t_1$ and $t_2$. 
Further discussion about how the MR-augmented classifier works is presented in the next section.

\subsection{Discussion}
\begin{figure}
	\centering
    \begin{subfigure}[b]{\textwidth}
        \includegraphics[width=\textwidth]{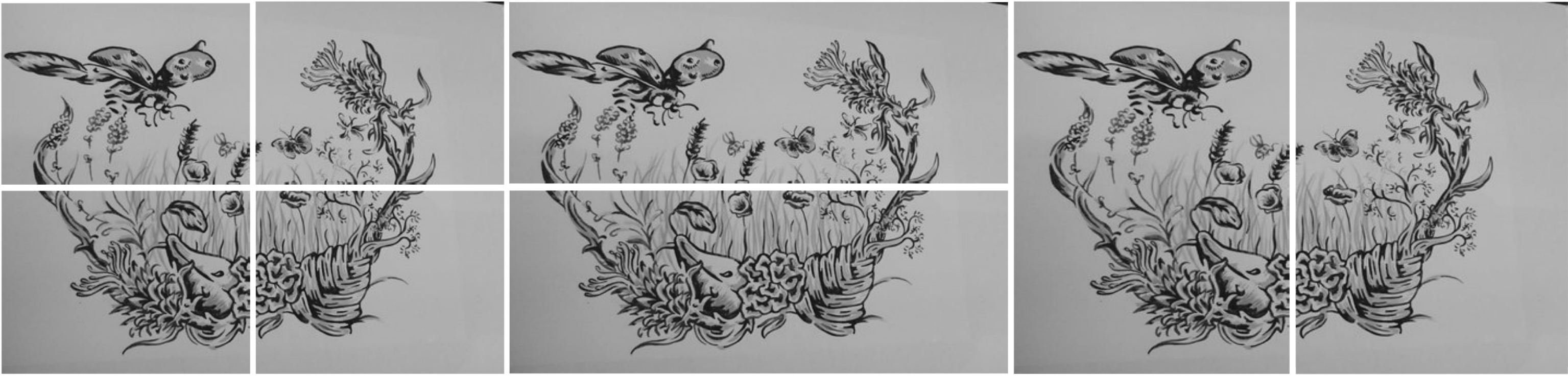}
    \end{subfigure} 
    \caption{Image blocks generated according to separation and occulsion.}
    \label{fig:imageBlocksByMRs}
\end{figure}

As explained in \sectionautorefname \ref{subsec:results}, the MR-augmented classifiers obtained better recall and precision results than the original classifiers (with approximately 10-15\% improvement). 
Recall and precision focus on the positive class (Artcode), with higher values indicating more confident and complete predictions of Artcodes, while some Artcode misclassifications by the original classifier were corrected by the MR-augmented classifier in the rectification stage. 
The decision as to whether or not to rectify was based on the $\rho$-value, as given in \equationautorefname \ref{eq:aggregatedLikelihood}, a measure of aggregated likelihood that an input image belongs to the Artcode class (\sectionautorefname \ref{subsec:rectification}).

As described in \sectionautorefname \ref{subsec:metaRelations}, the Separation and Occlusion MRs are based on the assumption that Artcode image blocks are more likely to be classified as {\em Artcode} than {\em non-Artcode}. 
The effectiveness of the two MRs was investigated by examining the prediction and rectification of image blocks for those images adjusted by the MR-augmented classifier (the red and blue points in \figureautorefname \ref{fig:rectificationIllustration}). 
The non-Artcodes that were incorrectly adjusted were the images that were very similar in topology to Artcodes (containing a number of connected regions), and had repeated geometrical structures, such as the 2nd and 4th images in \figureautorefname\ref{fig:nonartcodesExamples}. 
Repeated structures enabled the separate image blocks to inherit more topological structure from the original image, making their internal structures similar to those of Artcodes.
Occlusion and separation sometimes strengthened their topological structure, because occlusion and separation may remove auxiliary structures such as background imagery.
Accordingly, the MR-augmented classifiers are sensitive to this kind of {\it Artcode-like} images (such as the 4th image in \figureautorefname\ref{fig:nonartcodesExamples})  
--- 
images that are topologically very similar to Artcodes
--- 
which may result in incorrect rectifications.

Likewise, if separation and occlusion completely break the topological structure, Artcodes would be incorrectly rectified as non-Artcode by the MR-augmented classifiers. Fortunately, Artcodes have a topological structure that includes a number of connected regions, and often include several repeated structures with the same topology (but different geometry). 
These two properties enable Artcode image blocks to very likely retain the original topology, even after separation and occlusion. 
An example is presented in \figureautorefname \ref{fig:imageBlocksByMRs} for illustration: The image (the 5th in \figureautorefname \ref{fig:artcodesExamples}) is split into eight blocks by intersecting with the eight separation and occlusion masks shown in \figureautorefname \ref{fig:separationOcclusionMasks}
---
the left four image blocks in \figureautorefname\ref{fig:imageBlocksByMRs} are from separation, and the right four are from occlusion. 
Almost all of these blocks retain a complete topological structure:
they remain relatively complete Artcodes. 
Therefore, the MR-augmented classifier, based on the aggregated probability ($\rho$-value) of image blocks belonging to the {\em Artcode} class, can accumulate more information about this Artcode image than the original classifier, thereby achieving better overall predictions.

The two MRs are based on fundamental image processing operations, with the underlying rationale being whether or not the image blocks are able to retain the original structure's properties after transformations. 
Artcodes, as topological markers enabling redundancy, naturally possess this property.
The conventional use of MRs in metamorphic testing draws on intrinsic properties of the SUT. 
Likewise, the MRs used in Artcode classification also make use of intrinsic characteristics of Artcodes and non-Artcodes. 
Because domain characteristics may differ from task to task, and the repeated structures used in our two identified MRs may not exist in some contexts, it is likely that these MRs may not be directly applicable in some other image classification tasks. Nevertheless, this study has shown that MRs do have the potential to be used in image classification tasks (or even more general machine learning tasks), especially for those tasks with distinctive structural properties among different categories of learning data.

\section{Conclusion}\label{sec:conclusion}
This paper has reported on an examination of two previously identified MRs to enhance image classification, using them not only to improve performance, but also to explore verification of the classifier. 
Considering the uncertainty of classification algorithms, the verification exploration involved four statistical tests: 
one-way ANOVA, t-test (for unequal variances), Kruskal-Wallis test, and Dunnett's test. 
An effective and efficient MR-augmented classifier that uses SVM as the classification method, Aug-SVM, was introduced, and was compared with the Aug-RF classifier. 
The paper also examined the MR-augmented classification framework \citep{xu2018enhancing}, and presented a method that could be applied to related image classification problems for verification and enhancement.

Our experimental studies showed the applicability of ANOVA in conjunction with t-test (for unequal variances), ANOVA\_ranks, and Dunnett's test to explore verification of the classifier based on the two MRs.
The improved performance was not affected by the chosen classification method, demonstrating the potential to apply MT theories and techniques to general machine learning applications. 
Among the four classifiers in this paper (Ori-RF, Aug-RF, Ori-SVM, and Aug-SVM), Aug-SVM obtained the best performance in terms of both the evaluation metrics, and the computational efficiency. 
The experimental results also showed the essential role of the two thresholds, $t_1$ and $t_2$, for tuning the MR-augmented classifier performance. 
In addition, a theoretical analysis and discussion about how the enhanced performance was achieved by the MR-augmented classifiers was presented.

Our future work will include further examination of other parameters, including the number of masks ($n$ and $m$) for the separation and occlusion, the values in the weight vector $\vec{w}$, and the values of thresholds $t_1$ and $t_2$. 
A potential approach for choosing suitable values of $t_1$ and $t_2$ will be to examine the relationship between the thresholds and the centroids of $\rho_{G_a}$ and $\rho_{G_n}$. 
Because the work presented here has only examined two straightforward image transformations for the MRs, exploring other possible MRs that draw from other transformations for general image classification tasks, will also form part of our future work.

Although the two MRs employed in this work were straightforward, the results are promising, and clearly demonstrate the feasibility of MRs being used to augment classifiers. 
In order to fully investigate this new research area, more theoretical and practical work needs to be conducted, including exploration of connections between MRs and data augmentation, and case studies to examine the application of MRs to other well-studied image classification tasks (such as face and object detection) and even more broad machine learning problems. 
The concept of verifying machine learning software (the classifier in this paper) using MRs is still in an early stage of development, and more effort is also needed in the future. 
The proposed verification exploration based on ANOVA, t-test (for unequal variances), ANOVA\_ranks, and Dunnett's test, attempts to use statistical analyses to test {\em probabilistic} algorithms such as classification models:
Further work is necessary to extend this approach to verification, and fully explore its applicability.

\section*{Acknowledgements}\label{sec:acknowledgements}
This work was supported in part by the National Natural Science Foundation of China, under grant no.~61872167, by the Australian Research Council’s Discovery Projects funding scheme (Project ID: DP210102447), and by a Western River entrepreneurship grant.

\bibliographystyle{plainnat}  
\bibliography{references-JSS}

\end{document}